\documentstyle[aps,prb,epsf,multicol,amsfonts]{revtex}

\begin{document}
\draft
\tighten
\title{Coulomb blockade of strongly coupled quantum dots studied
via bosonization of a channel with a finite barrier}
\author{John M. Golden and Bertrand I. Halperin}
\address{Department of Physics, Harvard University, Cambridge, MA
  02138}

\date{22 August 2001}

\maketitle

\begin{abstract}
\noindent \ \
A pair of quantum dots, coupled to each other through a
point contact, 
can exhibit Coulomb blockade effects that
reflect the presence of an oscillatory term in the dots' total
energy whose value depends on whether the total number of 
electrons on the dots is even or odd.  The effective energy 
associated with this even-odd 
alternation is reduced, relative to the bare Coulomb blockade 
energy $U_{\rho}$ for uncoupled dots, by a factor $(1-f)$ 
that decreases as the 
interdot coupling is increased.  When the
transmission coefficient for interdot electronic motion
is independent of energy and is the same for all channels
within the point contact (which are assumed uncoupled), 
the factor $(1-f)$ takes on a universal value determined solely by
the number of channels $N_{\text{ch}}$ 
and the dimensionless conductance $g$ of each individual
channel.  When an individual
channel is fully opened (the limit $g \rightarrow 1$), 
the factor $(1-f)$ goes to zero. 

When the interdot transmission coefficient varies over energy
scales of the size of the bare Coulomb blockade energy $U_{\rho}$,
there are corrections to this universal behavior.  Here we
consider a model in which the point contact is described by
a single orbital channel containing a parabolic barrier potential,
with $\omega_{P}$ being the harmonic oscillator frequency associated
with the inverted parabolic well.
We calculate the leading correction to the factor $(1-f)$ for
$N_{\text{ch}} = 1$ (spin-split) and $N_{\text{ch}} = 2$ 
(spin-degenerate) point contacts, in the limit where $g$ is very
close to $1$ and the ratio $2 \pi U_{\rho}/\hbar \omega_{P}$ is
not much greater than $1$.
Calculating via a generalization of the bosonization technique
previously applied in the case of a zero-thickness barrier,
we find that for a given value of $g$, the value of $(1-f)$ is
increased relative to its value for a zero-thickness barrier, but
the absolute value of the increase is small in the region
where our calculations apply.
\end{abstract}

\pacs{PACS: 73.23.Hk,73.63.Kv,71.10.Pm,72.10.-d}

\begin{multicols}{2}

\narrowtext

\section{Introduction}

In recent years, there have been a number of theoretical and
experimental studies of the manner in which Coulomb blockade
effects in a metallic particle or semiconductor quantum dot
disappear when the conducting island becomes more electrically
connected to its environment.~\cite{Waugh,Crouch,Livermore,%
Flensberg,Matveev2,Molen,Matveev34,Golden1,Golden2,Golden3,Liu}
Here we focus on a system in which two symmetric quantum dots 
are defined by applying negative voltages to
gate electrodes that lie
on the surface of a semiconductor heterostructure
above a two-dimensional electron
gas (2DEG).  We assume that the dots are joined by a quantum point
contact containing a single orbital channel that is almost perfectly
transmitting at the Fermi energy, but that the dots are isolated
from their respective leads by comparatively large tunnel barriers.
In this geometry, information about the Coulomb blockade energy
$U_{\rho}$ involved in the transfer of electrons from one dot to
the other can be obtained by observing the positions of Coulomb
blockade peaks in the conductance across the entire system (from
one lead to another) when that conductance is plotted as a function
of voltages on gates coupled to each of the 
dots.~\cite{Waugh,Crouch,Livermore}

Previous analyses of the disappearance of the Coulomb blockade 
for two-dot systems~\cite{Note1} containing such a partially 
open point contact have characterized the contact by
a number $N_{\text{ch}}$ of degenerate one-dimensional (1D) 
channels and by the
dimensionless conductance $g$ of each channel 
(where $0 \leq g \leq 1$).~\cite{Molen,Matveev34,Golden1,Golden2}
This characterization is complete in the limit where the electronic
transmission amplitude through the contact is independent of
energy for electron energies that differ from the Fermi energy
by no more than an amount comparable to $U_{\rho}$.  
The energy independence of the transmission amplitude means that
this characterization corresponds to a potential barrier 
which is sufficiently thin that it can be modeled as a 
delta function.  Because
the Coulomb blockade energy $U_{\rho}$ is much smaller than the
Fermi energy $E_{F}$, working in this ``delta-function barrier 
limit'' can yield good results even though
the product of the barrier width and the Fermi wave vector $k_{F}$
is generally much greater than $1$ (with $k_{F}$ being the value of
the Fermi wave vector in the 2DEG far from the barrier).

Nevertheless, it seems important to investigate further 
the consequences of relaxing the assumption of a delta-function
barrier limit.  For one thing, it is possible to generate wider 
barriers using an appropriate gate geometry, and one would like to 
understand at what point the delta-function-limit calculations 
break down.  Secondly, estimating the corrections due to a barrier's
finite thickness provides a valuable check on the 
delta-function-limit results.

Another reason for interest is that recent
experiments on transmission through quantum point contacts have
shown unexpected structure (e.g., an apparent conductance
plateau near $0.7 (2 e^2 /h)$ at intermediate 
temperatures~\cite{Thomas})
whose origin is only poorly understood.  Such results suggest
that transmission through a point contact may have a nontrivial
energy dependence, and such a dependence could well have an
effect on the breakdown of the Coulomb blockade in the
coupled-dot geometry.  With the aim of explaining such effects 
in mind, it is a helpful first step to study theoretically 
the effects
of energy dependences in a simpler situation where the barrier
potential is known and many-body effects are reduced to a 
bare minimum.  In this vein we
consider here a model in which the two dots are separated by a
parabolic barrier of nonzero width and the electron-electron
interaction is taken to be constant for any two electrons
located anywhere on the same quantum dot.

This model presents challenges for ``bosonization
techniques'' that characterize point-contact constrictions 
as one-dimensional 
fermionic seas whose low-energy degrees of freedom can be 
expressed in terms of bosonic density and phase 
variables.~\cite{Emery,Heidenreich,Haldane,Fradkin,Schulz}  
In such ``bosonized'' models, the behavior of incompletely 
opened channels is commonly studied by introducing a 
zero-width delta-function barrier at a specific point in the 
one-dimensional sea.~\cite{Flensberg,Matveev2,Kane,Lal}  
By way of contrast, 
this paper seeks to show how a bosonized model of a 
one-dimensional
system can describe the single-particle effects of
replacing a delta-function barrier with a barrier that
more realistically corresponds to a finite-length 
constriction---i.e., a barrier of nonzero width and therefore
nontrivial single-particle transmission properties. 

Before proceeding, we should describe more fully what
is already understood about systems of two symmetric
quantum dots connected by a single orbital channel 
containing degenerate spin modes.  In such systems,
an energy scale $U_{\rho}$ characterizes the
energy cost of moving electronic charges between 
different segments of the system.  Moreover, the system's
Coulomb blockade behavior reflects the presence of an
energy term proportional to $U_{\rho}$ 
that oscillates between one value held when
the total number of electrons on the dots is even and
another value held when the total electron number is 
odd.~\cite{Waugh,Golden1}  

Previous work, both experimental
and theoretical, has shown that as the conductance of the
interdot point contact is increased, the energy scale
associated with this even-odd alternation is reduced by
a factor of $(1-f)$, where $f$ goes to $1$ when the
channel is fully open.  When the transmission coefficient
for electronic motion within this channel is independent
of energy (i.e., the delta-function barrier limit), 
the factor
$(1-f)$ takes on a universal value determined solely by
the number of point contact channels $N_{\text{ch}}$ 
(the channels being assumed
degenerate and uncoupled) and the dimensionless 
conductance $g$ of each individual channel (If $G$ is the
total conductance of the point contact, 
$g = G/N_{\text{ch}}(e^2/h)$.).  We refer to the value
$f$ in the factor $(1-f)$ as the {\em fractional peak
splitting} because it has been measured by observing
the relative separation of conductance peaks in a 
series of Coulomb blockade experiments.~\cite{Waugh,%
Crouch,Livermore,Golden1}

This paper studies the corrections to $f$, and therefore
to $(1-f)$, that result when the assumption of a delta-function
barrier limit is relaxed---in other words, when the 
interdot barrier is more realistically treated as having a
finite height and nonzero width.  A prior study of the
effects of a nonzero-width barrier concentrated on the
limit of weakly coupled dots (the limit 
$g \rightarrow 0$).~\cite{Golden3}
This study revealed
that, if the one-dimensional channels between two dots,
or between a dot and a lead, contain a tunneling barrier of 
finite height $V_{0}$ (in energy units) and of
nonzero width $\xi$, the behavior of such systems 
is responsive to another energy
scale, $W$, that characterizes the energy range
(for electrons incident on the barrier)
over which the probability of transmission
through the barrier varies substantially.~\cite{Golden3}
For relatively large and shallow dots, such as those
that have been constructed in
GaAs/AlGaAs heterostructures at low
temperatures,~\cite{Waugh,Crouch,Livermore}
the energies $U_{\rho}$ and $W$ 
tend to be much smaller than the Fermi energy $E_{F}$
but much larger than both the single-particle
level spacing $\delta$ and the
thermal energy $k_{\text{B}} T$.  As a result,
such systems are characterized by the following
hierarchy of energy scales:~\cite{Golden3}
\begin{equation}
k_{\text{B}} T, \delta
 \ll U_{\rho} \lesssim W \ll E_{F} \, .
\label{eq:hier}
\end{equation}
Because $W$ is comparable in size to the intermediate 
energy scale 
$U_{\rho}$ (which acts as a kind of ``excitation 
energy scale'' with respect to independent-particle
energies), effects from the barrier's finite size can be 
significant and deserve investigation before they are 
confidently discarded.

Our prior study of the weak-coupling limit where 
$g \ll 1$ found that the finiteness of the barrier leads 
to an upward correction to the universal  $f$-versus-$g$
curve for a delta-function barrier.  In other words,
for a given {\it small} 
value of $g$, the fractional peak
splitting $f$ is enhanced relative to its 
value for a delta-function barrier.~\cite{Golden3}
This enhancement
occurs because, in a channel containing a barrier with
a finite energy height (as opposed to a delta-function
barrier), electrons can tunnel from one dot to another
through largely unreflected states that have
single-particle energies greater than the energy that
corresponds to the finite barrier's peak.~\cite{Golden3}  

In order to study the correction to the universal
$f$-versus-$g$ curve, in our prior work we made 
several basic assumptions
about the system of two symmetric dots connected 
by a single orbital channel.
First, we assumed that the electrons that enter the
point contact between the dots are in the lowest
energy eigenstate for motion perpendicular to the
channel and can therefore be
described by a one-dimensional (1D) Schr\"{o}dinger
equation with an effective barrier potential $V(x)$.
We assumed that $V(x)$ could be treated as parabolic,
with a harmonic oscillator frequency $\omega_{P}$ 
associated with the inverted parabolic well.  (In
this case, the transmission energy scale $W$ 
is given by $\hbar \omega_{P}/2 \pi$.) 
In addition, we assumed that, despite the
nonzero width of the barrier, the essential
nature of electron-electron
interactions was still accurately represented
by the standard Coulomb blockade model, in which the 
interaction energy of a quantum dot is analogous
to that of a classical capacitor---i.e.,
proportional to the charge on the dot squared.
Having made these assumptions, we found
that, in the limit $g \rightarrow 0$, the leading 
behavior of the enhancement to $f$
is roughly proportional to
$N_{\text{ch}} (2 \pi U_{\rho}/\hbar \omega_{P})/|\ln{g}|$.
For the experimental systems with which we are concerned,
where $N_{\text{ch}}$ is $1$ or $2$ and where
$2 \pi U_{\rho}/\hbar \omega_{P} \simeq 1$, the 
size of the enhancement is
relatively small.~\cite{Golden3}

This paper extends study of such finite-barrier corrections 
to the strong-coupling limit where $(1-g) \ll 1$.
In doing so, we repeat the two assumptions 
described above.  The assumption
of a parabolic form for the imposed barrier potential 
is likely as good as 
before.~\cite{Lal,Buttiker}  However, our
simplistic treatment of the electron-electron 
interactions is likely 
less firm, given that the point contact is now largely
open and therefore much more likely to be occupied
by electrons interacting in a way not reflected by the
standard capacitive Coulomb blockade model.  
Nonetheless, because the barrier region is
still small in relation to the larger conducting basins
of the two dots, and because the effects from a combination
of the Coulomb blockade model and the interdot barrier's
finite size are themselves sufficiently interesting and complex,
in this initial stab at the problem of finite barriers 
we ignore the effects of interactions specific
to the point contact, with the understanding that 
separate efforts to understand those effects
should follow. 

Thus, the basic parameters for our strong-coupling study
are the same as those for the weak-coupling limit.
The $N_{\text{ch}}$ degenerate spin 
channels are assumed to contain effectively identical
barriers, and the interdot barrier can be modeled 
as fully parabolic
so long as we are concerned only with energies within
a restricted range about the barrier peak.
We therefore describe the single interdot 
orbital channel as containing the potential:
\begin{eqnarray}
V(x) & = & V_{0} \, (1 - x^2/2\xi^2) \ \ \ \ \ \ \ \ 
\text{for} \ \ |x| < \sqrt{2} \, \xi
 \nonumber \\
V(x) & = & 0 \ \ \ \ \ \ \ \ \ \ \ \ \ \ \ \ \ \ \ \ \ \ \ \
\ \  \text{for} \ \ |x| \geq \sqrt{2} \, \xi \, \, .
\label{eq:Vfunc}
\end{eqnarray}
See Figure~1.
As we will later see, our ultimate results are 
independent of the potential's details away from the
barrier peak.  Key aspects of the barrier potential are
the ``depth'' ($E_{F} - V_{0}$)
of its peak relative to the Fermi surface 
and the harmonic oscillator frequency
$\omega_{P} = \sqrt{V_{0}/m \xi^{2}}$
associated with the inverted parabolic well.~\cite{Golden3}
(Note that 
$\hbar \omega_{P}/E_{F} \approx \sqrt{2} \, (k_F \xi)^{-1}$,
which is assumed to be much less than $1$.)

We use the single-particle eigenfunctions 
that correspond to the above potential
to calculate corrections to $f$.  To do so,
we incorporate these eigenfunctions in a 
generalization of a bosonization approach~\cite{Emery,%
Heidenreich,Haldane,Fradkin,Schulz,Shankar} that was
used earlier in the delta-function barrier
limit.~\cite{Flensberg,Matveev2,Molen,Matveev34,%
Golden1,Golden2}
We first find the change in the two-dot system's 
ground state energy as the value of $(1-g)$ is increased
from zero.  This energy shift $\Delta(\rho)$ is
a function of $(1-g)$ and another dimensionless quantity $\rho$,
which represents a linear combination of the gate voltages 
applied to the two dots.~\cite{Golden1,Golden2}  When the total
number of electrons on the two dots is even, the 
fractional peak splitting $f$ is given by the
following formula:
\begin{equation}
f = 1 - \frac{\Delta(1) - \Delta(0)}{U_{\rho}/4} \, .
\label{eq:fgeneral}
\end{equation}
By applying this formula, 
we find that, consistent with our earlier 
conjecture,~\cite{Golden3} for both $N_{\text{ch}}=1$ and
$N_{\text{ch}}=2$ the leading finite-barrier
corrections {\it decrease} the value of $f$ observed 
for a given value of $g$ when $g$ is close to $1$.
For $2 \pi U_{\rho}/\hbar \omega_{P} \lesssim 2$ and
$(1-g) \ll 1$, the
magnitude of the decrease is proportional to
$(2 \pi U_{\rho}/\hbar \omega_{P})/|\ln{(1-g)}|$ for
$N_{\text{ch}} = 1$ and to
$(2 \pi U_{\rho}/\hbar \omega_{P}) \, \sqrt{1-g}
\, \, \{ 1 - |\ln{(1-g)}|^{-1} \}$ for
$N_{\text{ch}} = 2$.

\begin{figure}
        \begin{center}
                \leavevmode
        \epsfxsize=0.9\hsize
        \epsffile{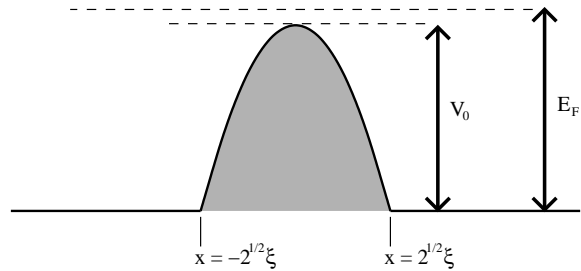}\\
        \end{center}
\caption{A schematic representation of the 1D model for a 
parabolic barrier between two quantum dots.  The half 
width of the barrier at half maximum
is $\xi$, and the potential energy at the barrier's peak
is $V_0$, which is less than, but comparable in size to, the 
Fermi energy $E_F$.}
\label{fig:1}
\end{figure}

In Sec.~II, we show how to use the bosonization approach to
derive an effective action that captures the Coulomb blockade
behavior of such a system for an arbitrary number
$N_{\text{ch}}$ of interdot
tunneling channels.
In Sec.~III, we use the effective action of Sec.~II
to solve for the leading $f$-versus-$g$ behavior
for $N_{\text{ch}} = 1$ and $g \simeq 1$.
In Sec.~IV, we
do the same for $N_{\text{ch}} = 2$.
In Sec.~V, we remark on the significance and
limitations of our results.

\section{Bosonization with a Barrier of Finite Width}

To calculate the corrections due to the interdot barrier's
nonzero width, we start with a fermionic formulation of 
our system
of two symmetric quantum dots connected by a single
orbital channel containing an arbitrary
number $N_{\text{ch}}$ of degenerate spin modes.
Since this fermionic problem is effectively
one-dimensional,~\cite{Golden3} we can apply to it
a variation of the standard
technique of bosonization.~\cite{Flensberg,Matveev2}
Having ``bosonized'' the problem, we integrate out
various bosonic degrees of freedom to
obtain a low-energy effective action involving
only a single set of scalar fields.

The fermionic formulation of the problem of
coupled quantum dots is quite 
straightforward.~\cite{Flensberg,%
Matveev2,Matveev34,Golden1,Golden2}
In a basis of non-interacting single-particle
eigenstates, the fermionic Hamiltonian consists of a
kinetic-energy part $H_{K}$ and a charging-energy
part $H_{C}$:
\begin{eqnarray}
H_{K} & = &
  \sum_{\sigma} \sum_{\zeta} \int \! \! dk \, \,
   \xi_{k} \, \, a^{\dagger}_{k \sigma \zeta} 
                        a_{k \sigma \zeta}
 \nonumber \\
H_{C} & = & U_{\rho} (\hat{n} - \rho/2)^{2} \ ,
\label{eq:ferm1}
\end{eqnarray}
where
$\xi_{k}$ is the energy of the single-particle eigenstate;
$\sigma$ is the channel index; $\zeta$ is the index for
states incident on the barrier from the left ($\zeta = 1$) and
the right ($\zeta = -1$), respectively; and $k_{F} + k$ is the
magnitude of the wave vector associated with motion along the
one-dimensional channel ($k_{F}$ is the Fermi wave vector and
$|k|$ is assumed to be smaller than $k_{F}$).  
The operator $\hat{n}$ measures the difference between the
number of electrons to the right of the barrier and
the number of electrons to the left
of the barrier.  If one lets $\hat{n}_{1}$ be the occupation
number for the left dot and $\hat{n}_{2}$ be the occupation
number for the right dot,
$\hat{n} = (\hat{n}_2 - \hat{n}_1)/2$.  In order to 
determine the value of $f$ via Eq.~\ref{eq:fgeneral}, 
we consider the case where the total number
of electrons on the two dots is even, so that $\hat{n}$ is
an operator with integer eigenvalues.  (When the total 
number of electrons is odd, $\hat{n}$ has half-integer 
eigenvalues.)

The operator $\hat{n}$ can be written in terms of fermionic
position operators $\psi_{\sigma}^{\dagger}(x)$ and
$\psi_{\sigma}(x)$ as follows:
\begin{equation}
\hat{n} = \frac{1}{2} \sum_{\sigma} \int \! \! dx \,
 [ \Theta(x) - \Theta(-x) ] \, \psi_{\sigma}^{\dagger}(x)
    \psi_{\sigma}(x) \, ,
\label{eq:nhat2}
\end{equation}
where $\psi_{\sigma}(x)$ consists of left-incident and
right-incident parts,
\begin{equation}
\psi_{\sigma}(x) = \psi_{\sigma 1}(x) + \psi_{\sigma, -1}(x) \, ,
\label{eq:psi}
\end{equation}
and where $\Theta(x)$ is a Heaviside step function centered
on the dividing line between the ``right'' and ``left'' sides
of the barrier at $x = 0$.

If one were bosonizing fermionic operators associated with a
basis of plane-wave states,
one could bosonize the above Hamiltonian by using the standard
formula:
\begin{equation}
\psi_{\sigma \zeta} (x) = \frac{1}{\sqrt{2 \pi \alpha}} \, 
        e^{i \zeta k_{F} x} \,
   e^{i  \sqrt{\pi} [ \zeta \theta_{\sigma}(x) \,
                                  + \,  \phi_{\sigma}(x) ]} \, \, ,
\label{eq:bosplane1}
\end{equation}
where $\theta_{\sigma}(x)$ and $\phi_{\sigma}(x)$ are scalar
boson fields and $\alpha$ is a quantity analogous to a lattice 
spacing that is used to impose an ultraviolet cutoff at energies 
approximating $\hbar v_{F}/\alpha$ on the bosonic degrees of 
freedom~\cite{Emery,Heidenreich,Haldane,Fradkin,Schulz,Shankar}
($\alpha$ is ultimately meant to be taken to zero---a limit
that can be reasonably taken when
$E_{F} \gg U_{\rho}$ and the approach to modeling the
double-dot system can be supposed to work
down to length scales $a$, where 
$\hbar v_{F}/a$ is much greater than the relevant 
excitation scale $U_{\rho}$).
The annihilation operators associated
with the plane-wave states
would then be given by the equation:
\begin{equation}
a_{k \sigma \zeta} = \frac{1}{2 \pi \sqrt{\alpha}}
   \int \! \! dx \, \, e^{- i \zeta k x} \, \,
   e^{i  \sqrt{\pi} [\zeta \theta_{\sigma}(x) \,
           + \, \phi_{\sigma}(x) ]} \, \, .
\label{eq:bosplane2}
\end{equation}
In terms of this plane-wave bosonization scheme, the operator
$\hat{n}$ could be simply expressed in terms of the boson
fields~\cite{Flensberg}:
\begin{equation}
\hat{n} = \frac{1}{\sqrt{\pi}} \sum_{\sigma} \theta_{\sigma} (0) \, .
\label{eq:nplane}
\end{equation}

Unfortunately, our problem is not so simple.  The
basis of single-particle states that
appears in Eq.~\ref{eq:ferm1} is not a
plane-wave basis.  The eigenstates are those of a
one-dimensional system containing a
barrier of finite height and nonzero width, and they
therefore include incident, reflected, and
transmitted parts.

Happily, the kinetic-energy part of the Hamiltonian
$H_{K}$ does not concern itself with such complications.
So long as the single-particle level spacing can be treated
as effectively zero,
we can proceed with regard to $H_{K}$ just as if the
single-particle states were plane waves.
Accordingly, if we focus on bosonizing $H_{K}$,
we can bosonize the annihilation operators in
Eq.~\ref{eq:ferm1} in the same way as in
Eq.~\ref{eq:bosplane2}.
Under this bosonization scheme,
the kinetic energy $H_{K}$ has the following
standard form:
\begin{equation}
H_{K} = \frac{v_{F}}{2} \sum_{\sigma} \int \! dx \, \left\{
 \left[ \partial_{x} \phi_{\sigma} (x) \right]^2
 + \left[ \partial_{x} \theta_{\sigma} (x) \right]^2 \right\}.
\label{eq:boskin}
\end{equation}

Our next task is to
find a proper bosonized form for the position
operators $\psi_{\sigma \zeta}(x)$ that appear in 
Eq.~\ref{eq:nhat2}.  Recalling that we already have
bosonized forms for the operators $a_{k \sigma \zeta}$,
we express the $\psi_{\sigma \zeta}(x)$ in bosonized form
by using the fact that
the position operators consist of linear combinations of the
operators $a_{k \sigma \zeta}$.  In particular, in the continuum
limit,
\begin{equation}
\psi_{\sigma \zeta}(x) =
        \int \! \! dk \, \, 
                Y_{k \zeta}(x) \, a_{k \sigma \zeta} \, ,
\label{eq:Ypsi}
\end{equation}
where the functions $Y_{k \zeta}(x)$ are the single-particle
eigenfunctions of the non-interacting one-dimensional 
system.~\cite{ExplanFunc}  For the potential described by
Eq.~\ref{eq:Vfunc}, we have derived these $Y_{k \zeta}(x)$
in prior work, using the fact that there are exact solutions
(parabolic cylinder functions) to the Schr\"{o}dinger equation
for a single particle in a parabolic potential.~\cite{Golden3}
Using the $Y_{k \zeta}(x)$, we
find that, to leading order in the perturbation
due to the barrier,
$\hat{n}$ equals $\hat{n}_0 + \delta \hat{n}$, where
\begin{eqnarray}
\hat{n}_0 & = & \frac{-1}{4 \pi i} \sum_{\sigma} \sum_{\zeta}
 \, \zeta \int \! \! dk_1 \, \int \! \! dk_2 \,
        \nonumber \\
 & & \hspace{0.2in} \times \,
 \left( \frac{1}{k_2 - k_1 + i \eta} \
      + \ \text{c.c.} \right)
    a_{k_2 \sigma \zeta}^{\dagger} a_{k_1 \sigma \zeta}
   \nonumber \\
\delta \hat{n} & = & \frac{1}{2 \pi} \sum_{\sigma}
 \int \! \! dk_1 \, \int \! \! dk_2 \, \,
 R(\epsilon_1, \epsilon_2)
\nonumber \\
 & & \hspace{0.2in} \times
 \, \left\{  \frac{e^{i [D(\epsilon_2) - D(\epsilon_1)]} }
      {k_2 - k_1 + i \eta}
\, a_{k_1 \sigma 1}^{\dagger} a_{k_2 \sigma, -1} \
 + \  \text{h.c.} \right\} \, ,
\label{eq:nparts}
\end{eqnarray}
where ``c.c.'' and ``h.c.'' stand, respectively, for the 
complex conjugate and hermitian conjugate of the preceding term,
and the following identities hold:
\begin{eqnarray}
\epsilon_{i} & = & \epsilon_{F} + v_{F} k_{i}/\omega_{P} \,
\nonumber \\
R(\epsilon_1, \epsilon_2) & = & \frac{e^{-\pi \epsilon_1} + 
e^{-\pi \epsilon_2}}{2 \sqrt{1 + e^{-2 \pi \epsilon_1}}
\, \sqrt{1 + e^{-2 \pi \epsilon_2}}}
\nonumber \\
D(\epsilon) & = & \frac{ \arg \Gamma(1/2 - i \epsilon) - \epsilon
        + \epsilon \ln{|\epsilon|} }{2} \, + \, C \, ,
\label{eq:parts}
\end{eqnarray}
with $C$ being a constant independent of $\epsilon$
and $\epsilon_{F} = \frac{E_{F} - V(0)}{\hbar \omega_{P}}$
being the value of $\epsilon$ at the Fermi energy.~\cite{Oops}

In the above equations, the
dimensionless variables 
$\epsilon_{i} = \frac{E_{i}-V(0)}{\hbar \omega_{P}}$ 
measure the
energy distance from the finite barrier's
peak.~\cite{Golden3}  The scale for these dimensionless
energy measures is the
harmonic oscillator energy $\hbar \omega_{P}$, and
in terms of these dimensionless measures
\begin{equation}
1-g = \frac{1}{1 + e^{2 \pi \epsilon_{F}}} \, \, .
\label{eq:oneminusg}
\end{equation}
The functions $R(\epsilon_1, \epsilon_2)$ and
$D(\epsilon_i)$ are associated with the 
single-particle transmission properties of the barrier.
Each additive term in $R(\epsilon_1, \epsilon_2)$ is a 
product of transmission and reflection amplitudes, 
reflecting the fact that the leading
contributions to $\delta \hat{n}$ come from
``overlaps'' between the transmitted part of an 
original right-mover and the reflected part of an
original left-mover and vice versa).
Meanwhile, $D(\epsilon_i)$ represents a phase associated
with the scattering of a single-particle wavefunction.
Because of the exponential growth (or damping) of
$R(\epsilon_1, \epsilon_2)$, we should not 
be surprised to find that 
$R(\epsilon_1, \epsilon_2)$ plays a starring role in the 
nature of our leading-order result---whereas, to
first approximation, contributions from
$D(\epsilon_i)$ will prove to be negligible.

Before we proceed further, we should make three remarks
about Eq.~\ref{eq:nparts}.  First, we should
note that derivation of the above equation for 
$\delta \hat{n}$ ignores contributions to
$\hat{n}_{0}$ from integrating over some interior
portion of the barrier region 
in which approximating the single-particle wavefunctions
by plane waves, or at least by WKB approximations to
plane waves, is no longer valid.  Hence---because
the exact eigenfunctions in the barrier region (parabolic
cylinder functions) do not ``explode'' in amplitude
as one goes further toward the center of the 
barrier~\cite{Miller}---as in our derivation of analogous 
identities in the limit of weakly coupled 
dots,~\cite{Golden3} we 
expect additive corrections of rough order $\xi$ to the 
Eq.~\ref{eq:nparts} cofactors 
$\frac{1}{k_2 - k_1 \pm i \eta}$.
Consequently, to avoid
substantial corrections to Eq.~\ref{eq:nparts}, we must
be able to restrict our attention to wave-vectors
$k_{i}$ (measured from the appropriate Fermi wave vector in 
one dimension) such that 
\begin{equation}
|k_{i}| \, \xi \lesssim 1.
\label{eq:waverestrict}
\end{equation} 
Because the exponentially variant function 
$R(\epsilon_1, \epsilon_2)$ strongly favors values of $k_i$ 
that correspond to $\epsilon_i$ near $0$,
the $k_i$ of most interest for calculating corrections to
the universal behavior are those for which 
$k_i \simeq -\omega_{P} \epsilon_{F}/v_{F}$.
The above restriction on $k_i$ can thus be seen to mean
that we need $\epsilon_{F} \lesssim \sqrt{2 E_{F}/V_{0}}$.
For systems (such as those this paper seeks to model) in
which $E_{F}/V_{0} \simeq 1$, the restriction then reduces 
(conservatively speaking) to $\epsilon_F \lesssim 1$ 
or, equivalently, $(1-g) \gtrsim 10^{-3}$.
Given that the relevant experiments with quantum dot 
systems~\cite{Waugh,Crouch,Livermore}
have generally not resolved $g$ to increments of $10^{-3}$, 
this constraint is not a serious one.

As a second point regarding Eq.~\ref{eq:nparts},
it should be noted that, for $\epsilon_F \lesssim 1$,
this intermediate result (and therefore those that
follow from it) is insensitive
to details of the barrier away from the barrier
peak (e.g., the slope discontinuities that occur in
our barrier potential at $x = \pm \sqrt{2} \, \xi$)
so long as $\xi$ is large compared to the inverse Fermi 
wave vector $k_{F}^{-1}$ (with the Fermi wavelength
itself being of the order of $2 \pi \hbar/\sqrt{2 m V_0}$, 
the wavelength
associated with the peak barrier energy, $V_0$).
The condition $\xi \gg k_{F}^{-1}$ is
typically true of the kinds of quantum dots 
that have motivated these 
investigations.~\cite{Waugh,Crouch,Livermore,Golden3}

A third remark about Eq.~\ref{eq:nparts} is 
that in deriving $\delta \hat{n}$, we have ignored
a correction involving terms of the form 
$(\cos [D(\epsilon_2) - D(\epsilon_1)] - 1)/(k_2 - k_1 \pm i \eta)$.
We ignored similar ``phase-based'' terms in 
Ref.~10.  The basic reason that we ignore these 
terms here derives from the fact that, because the 
characteristic excitation energy
$U_{\rho}$ is less than $\hbar \omega_{P}$, 
the variation of the numerator of these terms
over any region of particular interest should be very small.
As a result, in the regions that we generally find to be most
important (those regions near the 
simple poles where $k_{2} = k_{1} \pm i \eta$ and
the magnitudes of other contributing terms tend toward 
infinity),
these phase-based terms go to zero.   
Thus, we have good reason to expect that the contribution from 
these phase-based terms is small compared to the contribution 
from the terms we keep.  Indeed, if we follow the calculation
of the lowest-order contribution from these terms further,
we would find that, by reasoning parallel to that used
in Sec.~III to perform various wave vector integrals,
these terms produce a contour integral with no contribution from
residues of the simple poles at $k_{2} = k_{1} \pm i \eta$ and
only negligible, higher-order contributions from
integration around other singularities and branch cuts.

Returning to Eq.~\ref{eq:nparts} and the process of 
determining the fractional
peak splitting, we observe that if the operators $a_{k \sigma \zeta}$ 
were annihilation operators for particles in
plane-wave states (e.g., if there were no interdot barrier), 
$\hat{n}$ would simply equal $\hat{n}_0$.
In other words, $\hat{n}_0$ bears the same relation to the bosonic 
fields associated with the barrier-state operators
$a_{k \sigma \zeta}$ that $\hat{n}$ bears to the bosonic fields 
associated with the similarly indexed plane-wave operators
of Eq.~\ref{eq:bosplane2}.  From
Eq.~\ref{eq:nplane}, it therefore follows that
\begin{equation}
\hat{n}_0 = \frac{1}{\sqrt{\pi}} \sum_{\sigma} \theta_{\sigma} (0) \, .
\label{eq:nbarrier}
\end{equation}

We do not have a similarly simple formula for $\delta \hat{n}$.
Substituting for the operators $a_{k \sigma \zeta}$, we find that
\begin{eqnarray}
\delta \hat{n} & = & \frac{1}{(2 \pi)^3 \alpha} \sum_{\sigma}
 \int \! \! dx_1 \, \int \! \! dx_2 \,
 \int \! \! dk_1 \, \int \! \! dk_2 \
 R(\epsilon_1, \epsilon_2)
\nonumber \\
 & & \hspace{0.0in} \times
 \, \left\{  \frac{e^{i [D(\epsilon_2) - D(\epsilon_1)]}
                    e^{i (k_1 x_1 + k_2 x_2)} }
                  {k_2 - k_1 + i \eta}
 e^{-i \sqrt{\pi} [ \theta_{\sigma} (x_1) + \phi_{\sigma} (x_1) ] }
        \right. \nonumber \\
 & & \hspace{0.2in} \left. \times \,
 e^{-i \sqrt{\pi} [ \theta_{\sigma} (x_2) -  \phi_{\sigma} (x_2) ] }
 \ \ \ + \ \ \  \text{h.c.} \right\}
\label{eq:deltn}
\end{eqnarray}

Having now obtained bosonized expressions for the separate terms
$\hat{n}_{0}$ and $\delta \hat{n}$, we express the Hamiltonian
as a whole in bosonized form:
\begin{equation}
H = H_{K} + H_{C}^{(0)} + H_{C}^{(1)} + H_{C}^{(2)} \, .
\label{eq:Hamparts}
\end{equation}
$H_{K}$ is given by Eq.~\ref{eq:boskin}, and the remaining
terms are as follows:
\begin{eqnarray}
H_{C}^{(0)} & = & U_{\rho} \, \left[ \frac{1}{\sqrt{\pi}}
 \sum_{\sigma} \theta_{\sigma} (0) - \rho/2 \right]^{2} \,
 \nonumber \\
H_{C}^{(1)} & = & U_{\rho} \, \, \delta \hat{n}
  \left[ \frac{1}{\sqrt{\pi}}
 \sum_{\sigma} \theta_{\sigma} (0) - \rho/2 \right]
 \, + \, \text{h.c.} \,
 \nonumber \\
H_{C}^{(2)} & = & U_{\rho} \, (\delta \hat{n})^2 \, .
\label{eq:bosHam1}
\end{eqnarray}

The $H_{C}^{(2)}$ term can be dropped from the leading-order
calculation because, to second-order in $\delta \hat{n}$ (which
due to $R(\epsilon_1,\epsilon_2)$ is
itself roughly proportional to $\sqrt{1-g}$ at the Fermi
surface), $H_{C}^{(2)}$ 
contributes nothing to the $\rho$-dependence of the
ground-state energy.  Recall from Eq.~\ref{eq:fgeneral} that 
it is the $\rho$-dependence
of the ground-state energy that provides the basis for
calculation of the fractional peak-splitting $f$.~\cite{Golden1,%
Golden2}

To simplify calculation of this $\rho$-dependence,
we shift the $\theta_{\sigma}$-fields by a term
linear in $\rho$: $\theta_{\sigma} (x) \rightarrow
\theta_{\sigma} (x) + \sqrt{\pi} \rho/2 N_{\text{ch}}$.
This transformation leaves us with a Hamiltonian in which
the $\rho$-dependence appears only in the perturbative
factors of $\delta \hat{n}$.  The transformed (and
truncated) Hamiltonian has the following form:
\begin{eqnarray}
H_{K} &  = & \frac{v_{F}}{2} \sum_{\sigma} \int \! \! dx \, \left\{
 \left[ \partial_{x} \phi_{\sigma} (x) \right]^2
 + \left[ \partial_{x} \theta_{\sigma} (x) \right]^2 \right\} \,
 \nonumber \\
H_{C}^{(0)} & = & \frac{U_{\rho}}{\pi} \left[
 \sum_{\sigma} \theta_{\sigma} (0) \right]^{2} \,
 \nonumber \\
H_{C}^{(1)} & = & \frac{U_{\rho}}{\sqrt{\pi}}
 \sum_{\sigma} \left[ \delta \hat{n} \, \theta_{\sigma}(0)
 \, + \,  \theta_{\sigma}(0) \, \delta \hat{n} \right] \, ,
\label{eq:bosHam2}
\end{eqnarray}
where $\delta \hat{n}$ is now given by the identity:
\begin{eqnarray}
\delta \hat{n} & = & \frac{1}{(2 \pi)^3 \alpha} \sum_{\sigma}
 \int \! dx_1 \, \int \! dx_2 \, \int \! dk_1 \, \int \! dk_2 \,
 \, R(\epsilon_1, \epsilon_2)
\nonumber \\
 & & \hspace{0.0in} \times
 \, \left\{  \frac{e^{i [D(\epsilon_2) - D(\epsilon_1)]}
   e^{i (k_1 x_1 + k_2 x_2)} }
         {k_2 - k_1 + i \eta} \,
 e^{-i \pi \rho/N_{\text{ch}} } \, \right. \nonumber \\
 & & \hspace{0.2in} \times \,
 e^{-i \sqrt{\pi} [ \theta_{\sigma} (x_1) + \phi_{\sigma} (x_1) ] } \,
  \nonumber \\
 & & \hspace{0.2in} \left. \times \,
 e^{-i \sqrt{\pi} [ \theta_{\sigma} (x_2) -  \phi_{\sigma} (x_2) ] } 
 \ \ \  + \ \ \ \text{h.c.} \right\}
\label{eq:nhat2B}
\end{eqnarray}

With Eq.~\ref{eq:nhat2B}, we have reached the end of the road
with regard to the Hamiltonian approach.
To progress further, we shift to an action-based, path-integral
formulation.  For ease of notation, we will henceforth assume
that we have chosen units in which
\begin{equation}
\hbar = 1 \,
\end{equation}
and drop $\hbar$ from subsequent intermediate calculations 
(although we resuscitate it in stating our final results).
Having made this choice of units, we proceed by integrating 
out various degrees of freedom: first, the $\phi$-field degrees 
of freedom, and second, the $\theta$-field
degrees of freedom away from $x = 0$.  The result is an
effective action in terms
of fields $\theta_{\sigma}^{(0)}(\tau)$ that are
equivalent to the $\theta$-field degrees of freedom at
$x = 0$.
This effective action, dependent only on the $x = 0$ degrees
of freedom for the original $\theta$-fields, is analogous,
though---significantly---not equivalent, to that
previously obtained for a system containing a 
delta-function barrier (see Refs.~1 and 2).

We do no more than outline the process of integrating out the 
$\phi$-fields and $\theta$-fields because the methodology for
doing so is fairly straightforward.  With regard to
integrating out the $\phi$-fields, a key point is that,
in the path-integral approach, a Lagrangian term linear in
$\partial_x \phi (x,\tau) \, \partial_{\tau} \theta (x,\tau)$
appears in the action (just as a $p \dot{x}$-term appears in
going from the Hamiltonian for a single particle to the
corresponding Lagrangian).  Having accounted for this term,
we eliminate the $\phi$-fields by performing a standard 
Gaussian integration.  The
result is an effective action entirely in terms of
$\theta$-fields.

Having obtained this effective action, we
introduce the field $\theta_{\sigma}^{(0)} (\tau)$
and the auxiliary fields
$\lambda_{\sigma} (\tau)$.  We use the $\lambda$-fields
to enforce the identity
$\theta_{\sigma}(0,\tau) = \theta_{\sigma}^{(0)} (\tau)$ by
adding to the action the following term:
\begin{equation}
S_{\lambda} = \sum_{\sigma} \, \int \! \! d\tau \, \,
  i \lambda_{\sigma} (\tau) \left[ \theta_{\sigma}^{(0)}(\tau)
  - \theta_{\sigma} (0,\tau) \right] \, .
\label{eq:deltfact}
\end{equation}
Having added $S_{\lambda}$ to the action,
we can substitute $\theta_{\sigma}^{(0)} (\tau)$ for
$\theta_{\sigma} (0,\tau)$ in $H_{C}^{(0)}$
(recall Eq.~\ref{eq:bosHam2}).

Our next steps are to integrate out the
$\theta_{\sigma} (x,\tau)$-fields and then the
$\lambda_{\sigma} (\tau)$-fields.  Once again,
only straightforward---albeit tedious---Gaussian integrations
are required.  To leading order in the perturbative
factor $\delta \hat{n}$, the result is an
effective action dependent only on the scalar fields
$\theta_{\sigma}^{(0)}(\tau)$ and their Fourier transforms
$\tilde{\theta}_{\sigma}^{(0)}(\omega)$:
\begin{eqnarray}
S_{\theta}^{K} & = & \sum_{\sigma} \int \! \!
 \frac{d\omega}{2 \pi} \  | \omega | \
 \tilde{\theta}_{\sigma}^{(0)} (\omega)
 \tilde{\theta}_{\sigma}^{(0)} (-\omega)
 \nonumber \\
S_{\theta}^{C} & = & \sum_{\sigma_1} \,
 \sum_{\sigma_2} \, \int \! \! \frac{d\omega}{2 \pi} \,
 \left( \frac{U_{\rho}}{\pi} \right) \,
  \tilde{\theta}_{\sigma_1}^{(0)} (\omega)
  \tilde{\theta}_{\sigma_2}^{(0)} (-\omega)
 \nonumber \\
S_{\theta}^{(1)} & = & \sum_{\sigma_1} \,
 \sum_{\sigma_2} \, \int \! \! d\tau \,
 \int \! \! dx_1 \, \int \! \! dx_2 \, \int \! \! dk_1 \,
 \int \! \! dk_2 \
        \nonumber \\
 & & 
 \hspace{0.2in} \times \,
 \frac{2 U_{\rho}}{\sqrt{\pi} (2\pi)^3} \,
 R(\epsilon_1, \epsilon_2)  \,  A(x_1, x_2)  \,
 \nonumber \\
 & & \hspace{0.2in} \times \,
  \left\{ e^{i \pi \rho/N_{\text{ch}}} \, \,
 \frac{e^{-i [D(\epsilon_2) - D(\epsilon_1)]}
   e^{-i (k_1 x_1 + k_2 x_2)} }
      {k_2 - k_1 - i \eta} \, \, \right.
        \nonumber \\
 & & \hspace{0.5in} \times \,
  \theta_{\sigma_2}^{(0)} (\tau) \, \,
  e^{-i \int \! d\omega \, h_2(x_1,x_2,\omega) 
    \, e^{-i \omega \tau} \, 
        \tilde{\theta}_{\sigma_1}^{(0)}(\omega) }
  \nonumber \\
 & & \left. \hspace{0.4in}  
        + \ \text{m.t.} \right\} \, ,
\label{eq:bosAct}
\end{eqnarray}
where the ``mirror term'' (m.t.) can be obtained by complex
conjugating both the factors that precede 
$\theta_{\sigma_2}^{(0)}(\tau)$
and the exponential prefactor $i$ that precedes the 
integral over $\omega$, and
where the functions
$A(x_1, x_2)$, $h_2(x_1,x_2,\omega)$,
$h_1(x_1,x_2,\omega)$, and
$h_0(x_1,x_2,\omega)$ are given by the following formulae:
\begin{eqnarray}
A(x_1, x_2) & = & \frac{1}{(|x_1| + |x_2| + \alpha)^{1/2}} \,
        \nonumber \\
 & & \times \, \frac{1}{
 (2|x_1| + \alpha)^{1/4} \, (2|x_2| + \alpha)^{1/4} }
\nonumber \\
 & & \hspace{0.0in} \times \,
 \left[ \frac{ \alpha (2|x_1| + \alpha)^{1/2}
  (2|x_2| + \alpha)^{1/2} }
  { (|x_1| + \alpha)(|x_2| + \alpha) } \right]^{1/2}
        \nonumber \\
 & & \hspace{0.0in}
  \times \,
 \left[ \frac{ (|x_1| + \alpha)(|x_2| + \alpha) }
  { \alpha (|x_1| + |x_2| + \alpha) }
  \right]^{ \text{sgn} (x_1) \text{sgn} (x_2) /2 } \
 \nonumber \\
h_2(x_1, x_2, \omega) & = & h_1(x_1,x_2,\omega)
                                - h_0(x_1,x_2,\omega) \,
 \nonumber \\
h_1(x_1, x_2, \omega) & = & \text{sgn}(\omega) \,
  \frac{\text{sgn} (x_2) \left[ 1 - e^{-|\omega x_2|/v_F} \right]
        }
        {2 \sqrt{\pi}} \
 \nonumber \\
 & & \hspace{0.3in} - \ \{ x_1 \leftrightarrow x_2 \}
        \nonumber \\
h_0(x_1, x_2, \omega) & = & 
  \frac{e^{-|\omega x_1|/v_F} \, + \, e^{-|\omega x_2|/v_F} }
        {2 \sqrt{\pi}} \ ,
\label{eq:Afunc}
\end{eqnarray}
where ``$\{ x_1 \leftrightarrow x_2 \}$'' is the same
as the preceding term, except with the roles of $x_1$ and
$x_2$ reversed.

The effective action of Eq.~\ref{eq:bosAct} is the end result
for this section.  As promised, this action depends only on
a single set of scalar fields, which are 
equivalent to the original $\theta$-fields at $x = 0$.
Sections~III and IV use this action to solve for the fractional
peak splitting $f$
in systems with one and two interdot
channels, respectively.  In both sections,
the general approach is to treat
$S_{\theta}^{(0)} = S_{\theta}^{K} + S_{\theta}^{C}$ as the
unperturbed action and to solve perturbatively in
$S_{\theta}^{(1)}$.

\section{Finite-Barrier Result for the One-Channel Problem}

Here we consider the single-channel case ($N_{\text{ch}} = 1$),
in which electrons of only one spin
have enough energy to penetrate into the point contact.
Starting with the unperturbed action
$S_{\theta}^{(0)} = S_{\theta}^{K} + S_{\theta}^{C}$,
it is not hard to find that
\begin{equation}
\langle \tilde{\theta}^{(0)} (\omega_1) \,
 \tilde{\theta}^{(0)} (\omega_2) \rangle
 = \frac{\pi}{|\omega_1| + U_{\rho}/\pi}
   \, \, \delta (\omega_1 + \omega_2) \ .
\label{eq:thet1}
\end{equation}
We use this identity to calculate the
first-order correction to the ground state energy:
$\Delta (\rho) = \langle S_{\theta}^{(1)} \rangle / \beta$.
In particular, we find the following:
\begin{eqnarray}
\Delta(\rho) & = &
\frac{U_{\rho}}{(2 \pi)^3 \, \sqrt{\pi}} \, 
 \int \! \! dx_1 \int \! \! dx_2 \int \! \! dk_1 \int \! \! dk_2
\, \,  R(\epsilon_1, \epsilon_2)
\nonumber \\
 & & \hspace{-0.3in} \times \,
 \frac{e^{C_{2}(x_1, x_2, 2 U_{\rho})/4} }
  {(|x_1| + |x_2| + \alpha)^{1 /2} \, (2|x_1| + \alpha)^{1 /4}
   \, (2|x_2| + \alpha)^{1 /4} }
 \nonumber \\
 & & \hspace{-0.3in} \times \, \,
 e^{- (\pi /2)  \int \! d\omega \,
     \left[ h_0(x_1,x_2,\omega) \right]^{2} \,
 \frac{e^{- \alpha |\omega|/v_{F}}}{|\omega| \, + \, U_{\rho}/\pi} }
 \, \nonumber \\
 & & \hspace{-0.3in} \times \,
\int \! d\omega \,
    h_0(x_1,x_2,\omega) \,
 \frac{e^{- \alpha |\omega|/v_{F}}}{|\omega| + U_{\rho}/\pi}
 \nonumber \\
 & & \hspace{-0.3in} \times \, \left[ i \, e^{i \pi \rho} \,
 \frac{e^{-i [D(\epsilon_2) - D(\epsilon_1)]}
   e^{-i (k_1 x_1 + k_2 x_2)} }
      {k_2 - k_1 - i \eta} \
  + \ \text{c.c.} \right] ,
\label{eq:Dmess}
\end{eqnarray}
where the function $C_2(x_1,x_2,2U_{\rho})$ is defined
below in Eq.~\ref{eq:C1}.

Before we apply standard techniques of 
complex analysis to reduce the integral over $k_2$ to
the contributions from the simple poles at
$k_2 = k_1 \pm i \eta$, we need to show that we can 
neglect the effects of the singularities in the
expression for $S_{\theta}^{(1)}$ that result from
the factors $R(\epsilon_1, \epsilon_2)$ and $D(\epsilon)$.
It is not too hard to discount the singular behavior at 
$\epsilon_2 = 0$ that results from the functional form of
$D(\epsilon)$ (see Eq.~\ref{eq:parts}).
We can rewrite $\epsilon \ln |\epsilon|$ as follows:
\begin{equation} 
\epsilon \ln |\epsilon| = 
\epsilon \ln{\epsilon} 
- i \pi \epsilon \, \tilde{\Theta} (- \epsilon) \, ,
\label{eq:logeps}
\end{equation}
where $\tilde{\Theta}(-\epsilon)$ is a sort of variant
of the Heaviside step function that, in the appropriate
limits, is $1$ for $\epsilon$ real and negative 
and is $0$ for $\epsilon$ real and positive. 
To allow for complex values of $\epsilon$, we can
define $\tilde{\Theta}(-\epsilon)$ by the formula
$\tilde{\Theta}(-\epsilon) = \lim_{\upsilon \rightarrow 0} 
\int_{\Gamma(\epsilon)} \, dz \, 
        \frac{\upsilon}{\pi (z^2 + \upsilon^2)}$ ,
where the path $\Gamma(\epsilon)$ in the complex plane
starts at $z = R$ (with $R$ being a large and
positive real number, ultimately taken to $\infty$) and then 
proceeds to the value $\epsilon$ by the shortest path that avoids
the singularities at $z = \pm i \upsilon$ and their accompanying
branch cuts.  In the limit 
$R \rightarrow \infty$, this definition yields the desired behavior
for real values of $\epsilon$.  On the other hand, it does 
leave $\tilde{\Theta}(-\epsilon)$ with two branch points
(at $\epsilon = \pm i \upsilon$), and to these we must add the
branch point of $\epsilon \ln{\epsilon}$ at $\epsilon = 0$.

Fortunately, it is not too hard to see that integrating around
these singularities and associated branch cuts leads
to contributions to the final result that are higher-order in
the expansion parameter 
$2 \pi U_{\rho}/\hbar \omega_{P}$ than the contributions
from the simple poles at $k_2 = k_1 \pm i \eta$.
Rough evaluation of the higher-order contributions 
indicates that they are indeed negligible in the regime 
where $2 \pi U_{\rho}/\hbar \omega_P \lesssim 2$, but
are probably not negligible for values of 
$2 \pi U_{\rho}/\hbar \omega_P$ approaching $10$.

Having discounted higher-order contributions from 
singularities other than the 
simple poles at $k_2 = k_1 \pm i \eta$, we can perform the $k_2$
integral with ease.  The integral over $k_1$ can then be done
after the resulting factor $R(\epsilon_1,\epsilon_1)$ is expanded
in powers of $e^{\pm \pi \epsilon_1}$ (and after it is noted that 
$e^{-i [ D(\epsilon_1) - D(\epsilon_1) ] } = 1$), and the
result is the following:
\begin{equation}
\Delta (\rho) = - \frac{2 U_{\rho}}{ (2 \pi)^2} \, \cos (\pi \rho)
   \, J_1 (\epsilon_{F} , U_{\rho}, \omega_P) \ ,
\label{eq:en1ch}
\end{equation}
where, once again, 
$\epsilon_{F}$ is the ``dimensionless Fermi energy'' relative to
the barrier peak, $\omega_P$ is the harmonic oscillator energy of
the parabolic barrier (recall the discussion of
Sec.~I), and
\begin{eqnarray}
J_1 (\epsilon_{F} , U_{\rho}, \omega_P) & = &
 \int_{-\infty}^{\infty} \! dx_1  \int_{-\infty}^{\infty} \! dx_2 \, \,
 \left[ \Theta (x_1) + \Theta (-x_2) \right] \,
        \nonumber \\
 & & \hspace{-0.2in} \times \,  
 e^{i \epsilon_F \omega_P (x_1 + x_2)/v_F} \,
        F_1(x_1, x_2, U_{\rho}, \omega_P)       \ .
\label{eq:J1}
\end{eqnarray}

As before, $\Theta(x)$ is the Heaviside step function
and $v_F$ is the Fermi velocity.  Furthermore, the function
$F_1 (x_1, x_2, U_{\rho}, \omega_P)$ has the formula:
\begin{eqnarray}
F_1(x_1, x_2, U_{\rho}, \omega_P) & = & 
\left( \frac{\omega_P}{2 \sqrt{\pi} \, v_F} \right)
 \sum_{n= 0}^{\infty}(-1)^n 
        \nonumber \\
 & & \hspace{-0.2in} \times \, 
\frac{2(2 n + 1)\pi}{(2 n +1)^2 \pi^2 + (x_1 + x_2)^2 \omega_P^2 /v_F^2}
 \nonumber \\
 & &
 \hspace{-0.2in} \times \,  \frac{ e^{C_2(x_1, x_2, U_{\rho})/4} }
        {(|x_1| + |x_2| + \alpha)^{1/2} } \,
        \nonumber \\
 & & \hspace{-0.2in} \times \, \frac{1}{
  (2 |x_1| + \alpha)^{1/4} \,  (2|x_2| + \alpha)^{1/4} }
\nonumber \\
 & & \hspace{-0.2in} \times \, \,
 e^{- (\pi /2) \int \! d\omega \, 
   \left[ h_0(x_1,x_2,\omega)  \right]^2
  \frac{e^{-\alpha |\omega|/v_F} }{|\omega| + U_{\rho}/\pi} }
        \nonumber \\
 & & \hspace{-0.2in}
 \times \,
 \int \! d\omega \, h_0(x_1,x_2,\omega) \,
  \frac{e^{-\alpha |\omega|/v_F} }{|\omega| + U_{\rho}/\pi}  \, ,
\label{eq:F1}
\end{eqnarray}
where
\begin{eqnarray}
C_2(x_1, x_2, U_{\rho}) & = & C_1(x_1, x_2, U_{\rho}) 
        + C_1(x_2, x_1, U_{\rho}) \,
  \nonumber \\
C_1(x_1, x_2, U_{\rho}) & = &
 - \int_{0}^{1} \! dz \, \frac{e^{- ( \alpha U_{\rho}/ \pi v_F)z} }{z}
        \nonumber \\
 & & \hspace{-0.4in} \times \, 
  \left[\left( 1 - e^{-(U_{\rho} |x_1|/ \pi v_F)z} \right)^2 \, \right.
 \nonumber \\
 & & \hspace{-0.3in}
      - \, \text{sgn}(x_1) \, \text{sgn}(x_2)
   \left( 1 - e^{-(U_{\rho} |x_2|/ \pi v_F)z} \right)
        \nonumber \\
 & & \hspace{0.1in} \left. \times \,
   \left( 1 - e^{-(U_{\rho} |x_1|/ \pi v_F)z} \right)
  \right]
  \nonumber \\
 & & \hspace{-0.6in} 
        - \, 2 \left( e^{U_{\rho} |x_1|/ \pi v_F} - 1 \right)
   \left( 1 - \text{sgn}(x_1) \, \text{sgn}(x_2) \right)
        \nonumber \\
 & & \hspace{-0.2in} \times
  \int_{1}^{\infty} \! dz \, \,
      \frac{e^{- (\alpha + |x_1|) (U_{\rho}/\pi v_F)z}}{z}
  \nonumber \\
 & & \hspace{-0.6in}
   + \, \left( e^{2 U_{\rho} |x_1|/ \pi v_F} - 1 \right)
        \nonumber \\
 & & \hspace{-0.2in} \times
     \int_{1}^{\infty} \! dz \, \,
   \frac{e^{- (\alpha + 2 |x_1|) (U_{\rho}/\pi v_F)z}}{z}
  \nonumber \\
 & & \hspace{-0.6in}
   - \, \text{sgn}(x_1) \, \text{sgn}(x_2)
   \left( e^{U_{\rho} (|x_1| + |x_2|)/ \pi v_F} - 1 \right)
        \nonumber \\
 & & \hspace{-0.2in} \times
     \int_{1}^{\infty} \! \! dz \, \,
       \frac{e^{- (\alpha + |x_1| + |x_2|) (U_{\rho}/\pi v_F)z}}{z} \ .
\label{eq:C1}
\end{eqnarray}

Although the equation for $\Delta(\rho)$ may not be
entirely transparent, it is not hard to confirm,
by taking the limit $\omega_{P} \rightarrow \infty$, 
that the result for $f$ 
that follows from Eqs.~\ref{eq:fgeneral} and 
\ref{eq:en1ch} agrees with that
previously derived for a delta-function 
barrier.~\cite{Matveev34,Golden1,Golden2}
The key is to recognize that the relation 
$\sqrt{1-g} =
e^{- \pi \epsilon_F}/\sqrt{1 + e^{-2\pi \epsilon_F}} \,$
means that that 
$J_1 (\epsilon_F, U_{\rho}, \omega_P)
\rightarrow 2 e^{\gamma} \sqrt{1-g}$
as $2 \pi U_{\rho}/\omega_P \rightarrow 0$, where
$\gamma$ is the Euler-Mascheroni constant 
($\gamma \simeq 0.577$).

The behavior for $2 \pi U_{\rho}/\hbar \omega_P \neq 0$ can be
found through numerical integration.  The results for
various values of
$2 \pi U_{\rho}/\hbar \omega_P$ are displayed in Figure~2.
As predicted in our study of the 
weak-coupling limit ($g \rightarrow 0$),~\cite{Golden3} for a 
given value of $g$ in the vicinity of $1$,
the fractional peak splitting $f$ is reduced relative to that for
a delta-function barrier.  This strong-coupling depression of
the $f$-versus-$g$ curve becomes greater for larger values
of $2 \pi U_{\rho}/\hbar \omega_P$.

\begin{figure}
        \begin{center}
                \leavevmode
        \epsfxsize=0.9\hsize
        \epsffile{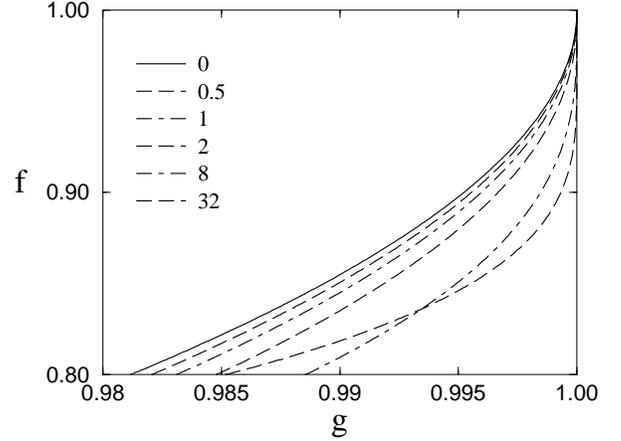}\\
        \end{center}
\caption{Plots of the leading $(1-g) \rightarrow 0$ behavior
of $f$, the fractional peak splitting, as a function of $g$,
the dimensionless interdot channel conductance, for a one-channel
connection between two quantum dots ($N_{\text{ch}} = 1$).
Each curve corresponds to a different value of the quantity
$2 \pi U_{\rho}/\hbar \omega_{P}$ (see legend to the left of the
curves).  The solid line is the result for an interdot barrier
that has effectively zero width
($2 \pi U_{\rho}/\hbar \omega_{P} = 0$).  The dashed and
dot-dashed curves show the leading $f$-versus-$g$ dependence for
finite barriers with $2 \pi U_{\rho}/\hbar \omega_{P}$ taking
on values from $0.5$ to $32$.  The curves are expected to be
substantially accurate at least for $(1-g)$ greater than or
equal to approximately $10^{-3}$ and for values of 
$2 \pi U_{\rho}/\hbar \omega_{P}$ not much larger than $2$.}
\label{fig:2}
\end{figure}

For $2 \pi U_{\rho}/\hbar \omega_P \simeq 1$, however, the
downward correction is quite small.  Since earlier
approximations of the barrier as a delta function were
designed to predict
the behavior of experimental systems in which
$2 \pi U_{\rho}/\hbar \omega_P \simeq 1$,~\cite{Waugh,%
Crouch,Livermore,Golden3} our calculation confirms that those
prior results were substantially correct.  Figure~2 gives a sense
of what the leading $2 \pi U_{\rho}/\hbar \omega_{P}$-dependent 
corrections look like for small and large
values of $2 \pi U_{\rho}/\hbar \omega_{P}$, but, because
of the approximations we have made, the results are expected
to be reliably accurate only 
for $2 \pi U_{\rho}/\hbar \omega_P \lesssim 2$.

One might wonder what analytic form the leading
$2 \pi U_{\rho}/\hbar \omega_P$-dependent corrections take.
In the weak-coupling limit ($g \rightarrow 0$), the 
finiteness of 
the barrier was shown to result in leading-order corrections
proportional to $(2 \pi U_{\rho}/\hbar \omega_{P})/|\ln g|$
for $2 \pi U_{\rho}/\hbar \omega_{P} \lesssim 2$.~\cite{Golden3}  
Corrections of this form arose from the fact that transitions into 
virtual states with energies near or above that of the 
barrier peak allow for relatively free interdot movement
(with transmission probabilities on the order of $1$).

By reasoning analogous (but in a sense opposite) to
that used in studying the weakly coupled system, 
we expect that, to leading order, 
the finite-barrier corrections in
the strong-coupling limit, where $(1-g) \ll 1$, are
dominated by the enhanced backscattering
of states near or below the barrier peak,
where reflection probabilities are on the order of $1$,
rather than on the order of $\sqrt{1-g}$ as is true
near the Fermi surface.  
To estimate these corrections from strongly backscattered
states, it is best to return to the
form of $\Delta(\rho)$ that appeared in 
Eq.~\ref{eq:Dmess}.

Examination of the $x$-dependent factors in Eq.~\ref{eq:Dmess}'s
integrand reveals that the primary contributions for the integral
come when the variables $x_{i}$ have values of 
rough magnitude $\pi \hbar v_{F}/2 U_{\rho}$.  In this range of 
$x_{i}$-values, the $k_{i}$-independent factors of the 
integrand (i.e., all the factors but $R(\epsilon_1,
\epsilon_2)$ and those enclosed in the final set of large
brackets) vary relatively slowly, only changing substantially
when one or both of the $x_{i}$ values changes by an
addend of order $\pi \hbar v_{F}/2 U_{\rho}$.  Consequently, if
the wave vectors $k_{i}$ have magnitudes greater than 
$2 U_{\rho}/\pi \hbar v_{F}$, the $k_{i}$-independent factors
can, to leading order, be treated as constants for
$x_{i} \sim \pi \hbar v_{F}/ 2 U_{\rho}$, and the 
$x_{i}$-dependence of the integrands in Eq.~\ref{eq:Dmess} 
can be assumed to be dominated by the factor 
$e^{\pm i(k_1 x_1 + k_2 x_2)}$.

The question then is whether the $k_i$ values that contribute
most substantially to the 
$2 \pi U_{\rho}/\hbar \omega_{P}$-dependence are large enough
(in magnitude) for this assumption to be justified.
As we have reasoned above, the leading contributions to
$2 \pi U_{\rho}/\hbar \omega_{P}$-dependence are
expected to come from $k_{i}$ such that energy of the 
corresponding eigenstate is near or below the barrier peak.  
In other words, the most significant
wave vectors are expected to be ones such that 
$E_{F} + \hbar v_{F} k_{i} \lesssim V_{0}$.  Consequently,
we expect these $k_{i}$ to be negative and to have magnitudes
approximating $k_0 = (\omega_{P}/v_{F})\epsilon_{F}$,
where (given Eq.~\ref{eq:oneminusg})
\begin{equation}
k_{0} \simeq \frac{\omega_{P}}{2 \pi v_{F}} |\ln (1-g)| \, .
\label{eq:kzero}
\end{equation}

As noted in Sec.~II, the anticipated importance of
negative wave vectors of approximate magnitude $k_{0}$ 
(which yield values for $\epsilon_i$ equal or near to
$0$) means that we must have $(1-g) \gtrsim 10^{-3}$ 
for our general approach to be valid.  Our desire to 
gain a more compact analytic approximation 
to the leading $2 \pi U_{\rho}/\hbar \omega_{P}$-dependent 
behavior now leads us to impose another requirement: 
$k_0 \gtrsim 2U_{\rho}/\pi \hbar v_{F}$.
From Eq.~\ref{eq:kzero}, this requirement means that we need 
$(1-g) \ll e^{- (2/\pi) (2 \pi U_{\rho}/\hbar \omega_{P})}$.
Combining this upper bound on $(1-g)$ with our previous 
lower bound, we find
that our general approach and the new assumption that 
we propose to make (that the 
$e^{\pm i(k_1 x_1 + k_2 x_2)}$-factors dominate the 
$x_i$-dependence of the integrands) are both valid only if 
$10^{-3} \lesssim (1-g) \lesssim e^{-(2/\pi) 
(2 \pi U_{\rho}/\hbar \omega_{P})}$.
The new assumption therefore allows a useful 
analytic approximation (i.e., an approximation with a 
significant range of $g$ values in which it is valid) 
when $2 \pi U_{\rho}/\hbar \omega_{P}$ is not much greater
than $2$.  For $2 \pi U_{\rho}/\hbar \omega_{P}$ values 
slightly over $10$, on the other hand, the assumption's utility 
``breaks down'' entirely, as the room between 
the upper and lower limits on $(1-g)$ vanishes.  Nonetheless, 
because we have already restricted our attention to 
$2 \pi U_{\rho}/\hbar \omega_{P} \lesssim 2$, this 
``breakdown'' is of no concern; for the values of 
$2 \pi U_{\rho}/\hbar \omega_{P}$ of real interest, our 
new assumption for the purpose of a compact 
analytic approximation imposes no more restraint than our
initial assumption that $(1-g) \ll 1$.

Having determined that we can assume 
$e^{\pm i(k_1 x_1 + k_2 x_2)}$-dominance, 
we can now proceed with estimating the corrections due
to wave vectors in the vicinity of $k_{0}$.
Approximation of the integrations based on the above assumptions
leads to the conclusion that, to first approximation, the 
correction to $\Delta(\rho)$ is proportional to 
$(2 \pi U_{\rho}/\hbar \omega_{P})/|\ln (1-g)|$.  
Thus, the leading finite-barrier 
correction to the fractional peak splitting $f$
may be roughly written as
$-c_{1} (2 \pi U_{\rho}/\hbar \omega_{P})/ |\ln{(1-g)}|$,
where $c_{1}$ is a small positive number.  The form of
this correction is parallel to that found in the weak-coupling
limit; in both cases, the leading 
$2 \pi U_{\rho}/\hbar \omega_{P}$-dependent terms are
proportional to 
$(2 \pi U_{\rho}/\hbar \omega_{P})/ |\ln{(\frac{1-g}{g})|}$.
The results displayed in Figure~2 bear out our approximation
to the analytic behavior of the leading strong-coupling
corrections.  The graphed corrections for 
$0 \leq (2 \pi U_{\rho}/\hbar \omega_{P}) \leq 2$
agree with the predicted analytic behavior, producing errors
of less than about 10\% for the behavior as a function of $(1-g)$
and of less than about 20\% for the behavior as a function of 
$2 \pi U_{\rho}/\hbar \omega_{P}$.  The graphed values suggest
that the value of $c_1$ is slightly less than $0.05$.

\section{Finite-Barrier Result for the Two-Channel Problem}

The downward shift of the $f$-versus-$g$ curve for one-channel
systems ($N_{\text{ch}} = 1$) is paralleled by 
a similar downward shift of the $f$-versus-$g$ curve for
two-channel systems ($N_{\text{ch}} = 2$).  However, the route
to the two-channel result is less straightforward than the already
somewhat tortuous route taken
for the single-channel problem, largely due to the fact that we now
have to deal with two $\theta_{\sigma}^{(0)}$-fields in
Eq.~\ref{eq:bosAct}, rather than just one.
The first step in dealing with this set of fields is
to transform them to a more manageable pair---the
``charge'' and ``spin''
fields $\theta_{C}$ and $\theta_{S}$,
respectively---which are linear combinations of the
$\theta_1$ and $\theta_2$ fields:
$\theta_{C}  =  \theta_{1}^{(0)} + \theta_{2}^{(0)}$
and $\theta_{S}  =  \theta_{1}^{(0)} - \theta_{2}^{(0)}$.
The bosonic action $S$ then consists of the sum of separate
charge and spin contributions to the unperturbed action
$S_{0}$, and a perturbative term $S_{1}$, which depends on
both the charge and the spin fields.  In particular,
$S_{0} = S_{0}^{(C)} + S_{0}^{(S)}$ and $S = S_{0} + S_{1}$,
where
\begin{eqnarray}
S_{0}^{(C)} & = &
 \int \! \!
 \frac{d \omega}{2 \pi} \,
 \left( \frac{| \omega |}{2} + \frac{U_{\rho}}{\pi} \right) \,
 \tilde{\theta}_{C} (\omega)
 \tilde{\theta}_{C} (-\omega)
 \nonumber \\
S_{0}^{(S)} & = &
 \int \! \!
 \frac{d \omega}{2 \pi} \, \, \frac{| \omega |}{2}  \, \,
 \tilde{\theta}_{S} (\omega)
 \tilde{\theta}_{S} (-\omega)
 \nonumber \\
S_{1} \  & = &
 \frac{2 U_{\rho}}{\sqrt{\pi} (2\pi)^3} \,
  \int \! \! d\tau
 \int \! \! dx_1 \int \! \! dx_2 \int \! \! dk_1
 \int \! \! dk_2 \, \, 
        \nonumber \\
 & & \hspace{-0.2in} \times \,
 R(\epsilon_1, \epsilon_2)  \,  A(x_1, x_2)  \,
        \nonumber \\
 & & \hspace{-0.2in} \times \, \left\{ e^{i \pi \rho/2} \,
 \frac{e^{-i [D(\epsilon_2) - D(\epsilon_1)]} \,
   e^{-i (k_1 x_1 + k_2 x_2)} }
      {k_2 - k_1 - i \eta} \right.
        \nonumber \\
 & & \hspace{0.1in} \times \,
  \theta_{C} (\tau) \, \,
  e^{-(i/2) \int \! d\omega \, h_2(x_1,x_2,\omega)
 \,  e^{-i \omega \tau} \, \tilde{\theta}_{C}(\omega) }
  \nonumber \\
 & & \hspace{-0.05in} \left.
   + \ \ \text{m.t.} \right\}
  \nonumber \\
 & & \hspace{-0.2in} \times \,
 \left\{ e^{-(i/2) \int \! d\omega \, h_2(x_1,x_2,\omega) \,
   e^{-i \omega \tau} \,
  \tilde{\theta}_{S}(\omega) }
  \ \ + \ \ \text{m.t.} \right\}
  \, ,
\label{eq:bosAct2}
\end{eqnarray}
where the ``mirror terms'' (m.t.) can be obtained in the
same way as for Eq.~\ref{eq:bosAct}.

After integrating out the high-energy charge fields,
one is left with an effective spin action
$S^{(S)} = S_{0}^{(S)} + S_{1}^{(S)}$,
where $S_{0}^{(S)}$ is unchanged from
Eq.~\ref{eq:bosAct2} and where
\begin{eqnarray}
S_{1}^{(S)} & = &
 \frac{ U_{\rho}}{(2 \pi)^3 \sqrt{\pi}} \, \int_{0}^{\beta} \! \! d\tau
 \int \! \! dx_1 \int \! \! dx_2 \int \! \! dk_1 \int \! \! dk_2
        \nonumber \\
 & & \hspace{-0.1in} \times
  \,  R(\epsilon_1, \epsilon_2) \, A(x_1, x_2) \,
 e^{C_{2}(x_1, x_2, 2 U_{\rho})/8}
 \nonumber \\
 & & \hspace{-0.1in}
 \times \,
 \left[ \frac{\alpha \, (2|x_1| + \alpha)^{1/2} \,
        (2|x_2| + \alpha)^{1/2}}
   {(|x_1| + \alpha) \, (|x_2| + \alpha)} \right]^{-1/4}
        \nonumber \\
 & & \hspace{-0.1in} \times \,
 \left[ \frac{(|x_2| + \alpha) \, (|x_1| + \alpha)}
   {\alpha \, (|x_1| + |x_2| + \alpha)}
  \right]^{-\text{sgn}(x_1) \text{sgn}(x_2)/4}
 \nonumber \\
 & & \hspace{-0.1in} \times \, e^{- (\pi/4)  \int \! d\omega \,
   \left[ h_0(x_1,x_2,\omega) \right]^{2}
 \frac{e^{- \alpha |\omega|/v_{F}}}{|\omega| \, + \, 2 U_{\rho}/\pi} }
        \nonumber \\
 & & \hspace{-0.1in}
 \times \, \int \! d\omega \,
        h_0(x_1,x_2,\omega) \, 
 \frac{e^{- \alpha |\omega|/v_{F}}}{|\omega| + 2 U_{\rho}/\pi}
 \nonumber \\
 & & \hspace{-0.1in} \times \, \left[ i \, e^{i \pi \rho/2} \,
 \frac{e^{-i [D(\epsilon_2) - D(\epsilon_1)]} \,
   e^{-i (k_1 x_1 + k_2 x_2)} }
 {k_2 - k_1 - i \eta} \
  + \ \text{c.c.} \right]
 \nonumber \\
 & & \hspace{-0.1in} \times \,
 \left[ e^{-(i/2)  \int \! d\omega \, h_2(x_1,x_2,\omega) \,
 e^{-i \omega \tau} \, \tilde{\theta}_{S}(\omega) }
 \ \ + \ \ \text{m.t.}
 \right]  \, ,
\label{SpinAct1}
\end{eqnarray}
where, as was done in the case of the $N_{\text{ch}} = 1$ problem, 
only the lowest-order perturbative term has
been retained in the effective action.

To exploit the even-odd combinations that occur in the
integrals over $\omega$, we rewrite the action in terms of
the following ``even'' and ``odd'' fields:
\begin{eqnarray}
\tilde{\theta}_{S}^{(E)} & = &
 e^{- i \omega \tau} \tilde{\theta}_{S} (\omega)
 + e^{i \omega \tau} \tilde{\theta}_{S} (-\omega)
\nonumber \\
\tilde{\theta}_{S}^{(O)} & = &
 i \left[ e^{- i \omega \tau} \tilde{\theta}_{S} (\omega)
 - e^{i \omega \tau} \tilde{\theta}_{S} (-\omega) \right] \, .
\label{eq:EOfields}
\end{eqnarray}
The nonperturbative term $S_{0}^{(S)}$ then takes the form
\begin{eqnarray}
S_{0}^{(S)} & = & \frac{1}{4} \int_{0}^{\infty} 
        \frac{d\omega}{2\pi} \, \omega \, 
        \left[ \tilde{\theta}_{S}^{(E)} (\omega,\tau) \,
                \tilde{\theta}_{S}^{(E)} (\omega,\tau) \,
        \right. \nonumber \\
 & & \hspace{0.2in} \left.
        + \,  \tilde{\theta}_{S}^{(O)} (\omega,\tau) \,
              \tilde{\theta}_{S}^{(O)} (\omega,\tau) \right] \, .
\label{eq:SzOE}
\end{eqnarray}  
After integrating out the $\tilde{\theta}_S^{(O)}$-fields, we
have a new effective action in terms of the
$\tilde{\theta}_S^{(E)}$-fields.

The ``even'' fields are not to be left alone, 
however.  In anticipation of ``refermionization,'' we insert into 
the effective action a set of dummy ``odd'' fields---free
fields that are entirely decoupled from the ``even'' fields. 
The nonperturbative, kinetic energy
terms of these new $\tilde{\theta}_{S}^{(O)}$-fields and
$\tilde{\theta}_{S}^{(E)}$-fields add to give a total
kinetic energy of the same form as that in Eq.~\ref{eq:SzOE}.  
This total is in turn rewritten in a form equivalent to that of
$S_{0}^{(S)}$ in Eq.~\ref{eq:bosAct2}.  Our newest, and
final, effective action then consists of the sum of this 
$S_{0}^{(S)}$-term and the leading perturbative term
$S_{1}^{\text{f}}$, which was derived from integrating out the
original ``odd'' fields:
\begin{eqnarray}
S_{1}^{\text{f}} & = &
 \frac{ U_{\rho}}{(2 \pi)^3 \sqrt{\pi}} \, 
        \int_{0}^{\beta} \! \! d\tau
 \int \! \! dx_1 \int \! \! dx_2 \int \! \! dk_1
        \int \! \! dk_2 \, \,
  \,
\nonumber \\
 & & \hspace{0.0in} \times \,
 \frac{R(\epsilon_1, \epsilon_2) \, e^{C_{2}(x_1, x_2, 2 U_{\rho})/8} }
  { (|x_1| + |x_2| + \alpha)^{1/2} \, (2|x_1| + \alpha)^{1/4}
   \, (2|x_2| + \alpha)^{1/4} }
 \nonumber \\
  & & \hspace{0.0in} \times \,  e^{- (\pi/4)  \int \! d\omega \,
     \left[ h_0(x_1,x_2,\omega) \right]^{2}
 \frac{e^{- \alpha |\omega|/v_{F}}}{|\omega| \, + \, 2 U_{\rho}/\pi} }
        \nonumber \\
 & & \hspace{0.0in}
  \times \, \int \! d\omega \,
     h_0(x_1,x_2,\omega) \,
 \frac{e^{- \alpha |\omega|/v_{F}}}{|\omega| + 2 U_{\rho}/\pi}
 \nonumber \\
 & & \hspace{0.0in} \times \left[ i \, e^{i \pi \rho/2} \,
 \frac{e^{-i [D(\epsilon_2) - D(\epsilon_1)]} \,
   e^{-i (k_1 x_1 + k_2 x_2)} }
      {k_2 - k_1 - i \eta}
  \ + \ \text{c.c.} \right]
        \nonumber \\
 & & \hspace{0.0in} \times \,
  \left[ e^{ (i/2) \int \! d\omega \,
        h_0(x_1,x_2,\omega) \,
 e^{-i \omega \tau} \, \tilde{\theta}_{S} (\omega) }
  \ +  \ \text{m.t.}
 \right]  \, .
\label{eq:SpinActfin}
\end{eqnarray}

As in prior work,~\cite{Matveev2,Golden2}
we ``refermionize'' the bosonic action 
$S = S_{0}^{(S)} + S_{1}^{\text{f}}$ by identifying
$\psi_{f} (0, \tau)$ as
a fermionic annihilation operator at the origin 
of a semi-infinite system, where
\begin{eqnarray}
\psi_{f} (0, \tau) & = & 
  \frac{ e^{- (i/2) \, \int \! d\omega \,
  h_0(x_1, x_2, \omega) \,
 e^{- i \omega \tau} \, \tilde{\theta}_{S} (\omega) }}
        {\sqrt{2\pi \, \tilde{\alpha} (x_1,x_2) } } \, ,
\label{eq:referm2}
\end{eqnarray}
where
\begin{equation}
\tilde{\alpha} (x_1,x_2) = 
(2|x_1| + \alpha)^{1/4}
(2|x_2| + \alpha)^{1 /4} (|x_1| + |x_2| + \alpha)^{1/2} .
\label{eq:referm2_support}
\end{equation}
This refermionization formula may look peculiar 
because of 
its use of a normalization factor and an ultraviolet cutoff 
(embedded in $h_0 (x_1,x_2,\omega)$) that depend on position 
variables, rather than mere constants 
(contrast the standard bosonic representation of fermionic 
position operators in Eq.~\ref{eq:bosplane1}). 
These position-dependent factors work, in tandem with
the scalar prefactor $(i/2)$ for the exponentiated integral,
to ensure the correct anticommutation relations for 
$\psi_{f} (0, \tau)$, and their use is expected to be 
unobjectionable
so long as for all $x_{i}$ of real interest the resulting
ultraviolet cutoff is high enough to capture the
behavior with which we are concerned.  

As in the 
single-channel problem, the $x_{i}$ of real interest
satisfy $|x_{i}| \sim \pi \hbar v_{F}/2 U_{\rho}$ (as can be seen
by doing the integrals over $k_{i}$ and obtaining a result
with factors analogous to those of 
$F_{1}(x_1,x_2,U_{\rho},\omega_{P})$ of Eq.~\ref{eq:F1}).
The ultraviolet cutoffs in Eq.~\ref{eq:referm2} therefore 
characteristically correspond to energies approximating
$2 U_{\rho}/\pi$---which is much greater than
$k_{B} T$, the characteristic energy for
the unperturbed spin degrees of freedom, and (as we can 
confirm once we rediscover the leading order behavior 
produced by $S_{1}^{\text{f}}$), much greater than the
energy of states characteristically brought into play by the 
perturbation $S_{1}^{\text{f}}$.  Indeed, we have already
incorporated an assumption that ultraviolet cutoffs 
approximating $2 U_{\rho}/\pi$ are valid because, in 
obtaining our effective action, we have only kept the leading
order terms from integrating out the charge degrees of
freedom---an approximation only expected to be good if
the spin-field states of concern have
excitation energies significantly less than 
$U_{\rho}$.~\cite{Matveev2}

Returning to Eq.~\ref{eq:SpinActfin}, we use
the identity $\sqrt{2\pi} \int_{0}^{\beta} \! \! d\tau
\left[ \psi_{f} (0, \tau) + \psi_{f}^{\dagger} (0, \tau) \right]
= \int \! \! dk \, (f_{k}^{\dagger} + f_{k})$
to find that our refermionization scheme produces a fermionic 
Hamiltonian
$H = H_{0} + H_{1}$ that consists of the following parts:
\begin{eqnarray}
H_{0} & = & \int \! \! dk \, \, \xi_{k} \, \,
        f_{k}^{\dagger} f_{k}\,
 \nonumber \\
H_{1} & = & Z(\epsilon_{F}, U_{\rho}, \rho)
 \int \! \! dk \, \, (f_{k}^{\dagger} + f_{k}) \, ,
\label{eq:refermHam}
\end{eqnarray}
where 
\begin{eqnarray}
Z(\epsilon_{F}, U_{\rho}, \omega_{P}, \rho) & = &
\frac{ U_{\rho}}{(2 \pi)^3 \sqrt{\pi}} \, 
 \int \! \! dx_1 \int \! \! dx_2 \int \! \! dk_1 \int \! \! dk_2
\, \, 
\nonumber \\
 & & \hspace{-1.0in} \times \,
\frac{ R(\epsilon_1, \epsilon_2) \, e^{C_{2}(x_1, x_2, 2 U_{\rho})/8} }
  {(|x_1| + |x_2| + \alpha)^{1 /4} \, (2|x_1| + \alpha)^{1 /8}
   \, (2|x_2| + \alpha)^{1 /8} }
 \nonumber \\
 & & \hspace{-1.0in} \times \, e^{- (\pi/4)  \int \! d\omega \,
    \left[ h_0(x_1,x_2,\omega) \right]^{2}
\frac{e^{- \alpha |\omega|/v_{F}}}{|\omega| \, + \, 2 U_{\rho}/\pi} }
        \nonumber \\
 & & \hspace{-1.0in}
 \times \, \int \! d\omega \,
     h_0(x_1,x_2,\omega) \,
 \frac{e^{- \alpha |\omega|/v_{F}}}{|\omega| + 2 U_{\rho}/\pi}
 \nonumber \\
 & & \hspace{-1.0in} \times \, \left[ i \, e^{i \pi \rho/2} \,
 \frac{e^{-i [D(\epsilon_2) - D(\epsilon_1)]}
   e^{-i (k_1 x_1 + k_2 x_2)} }
      {k_2 - k_1 - i \eta} \
  + \ \text{c.c.} \right] \, .
\label{eq:Zmess}
\end{eqnarray}

From prior work on the effects of a delta-function 
barrier, we know how to solve for the fractional peak 
splitting that such a Hamiltonian 
produces.~\cite{Matveev34,Golden1,Golden2} 
The only difference from the delta-function barrier Hamiltonian
that we encountered before is that a complicated prefactor
$Z(\epsilon_{F}, U_{\rho}, \omega_{P}, \rho)$
replaces the simpler delta-function barrier prefactor
$Z_{\infty}(\epsilon_F, U_{\rho}, \rho) = \cos (\pi \rho/2) \,
\sqrt{1-g} \, \sqrt{2 e^{\gamma} \hbar v_{F}U_{\rho}/\pi^3}$,
where $\gamma$ is the Euler-Mascheroni constant 
($\gamma \simeq 0.577$).~\cite{Matveev34,Golden1,Golden2}
Thus, to find the leading behavior of the fractional
peak splitting as a function of $(1-g)$ and 
$2 \pi U_{\rho}/\hbar \omega_{P}$, we can simply substitute
$Z(\epsilon_{F}, U_{\rho}, \omega_{P}, \rho)$ for
$Z_{\infty}(\epsilon_F, U_{\rho}, \rho)$ in the results
previously obtained for a delta-function barrier.

As for one-channel systems,
we ultimately resort to numerical integration
to solve for the fractional peak splitting when the
barrier has a nonzero width.
Needless to say, it is gratifying that such
numerical calculation confirms that, at least through
five significant digits,
the prefactor $Z(\epsilon_{F}, U_{\rho}, \omega_{P}, \rho)$ converges
to the delta-function quantity $Z_{\infty}(\epsilon_F, U_{\rho}, \rho)$
in the limit $2 \pi U_{\rho}/\hbar \omega_{P} \rightarrow 0$.
Consequently, in the limit of a narrow barrier, we recover
the same result as for a delta-function
barrier~\cite{Matveev34,Golden1,Golden2}---a good confirmation
both of the robustness of our earlier results and of our success
in wending through the complications created by the initial
assumption of a nonzero-width barrier.

The approach to performing the quadruple integral
of Eq.~\ref{eq:Zmess} is very similar to that used to perform
the analogous quadruple integral for $N_{\text{ch}} = 1$ 
and therefore will not be described in depth.
We first integrate over $k_1$ and $k_2$ by
techniques of complex analysis similar to those
used for $N_{\text{ch}} = 1$, once again 
focusing on the simple poles at $k_{2} = k_{1} \pm i\eta$, 
with the understanding that this imposes a constraint that
$2 \pi U_{\rho}/ \hbar \omega_{P} \lesssim 2$. 
For various values of 
$2 \pi U_{\rho}/\hbar \omega_{P}$, we then do the 
remaining integrals over $x_1$ and $x_2$ numerically.
The results are shown in Figure~3.
As for one-channel systems, we see that
as $2 \pi U_{\rho}/\hbar \omega_P$ increases,
the strong-coupling end of the $f$-versus-$g$ curve
shifts downward from the zero-width result.
Once again, the corrections for the experimentally
realized values of $2 \pi U_{\rho}/\hbar \omega_{P}
\simeq 1$ are small, a fact which confirms that the
previous assumption of a delta-function 
barrier~\cite{Matveev34,Golden1,Golden2,Golden3}
was substantially justified, at least so long as
interaction effects peculiar to the barrier region can
be ignored.

\begin{figure}
        \begin{center}
                \leavevmode
        \epsfxsize=0.9\hsize
        \epsffile{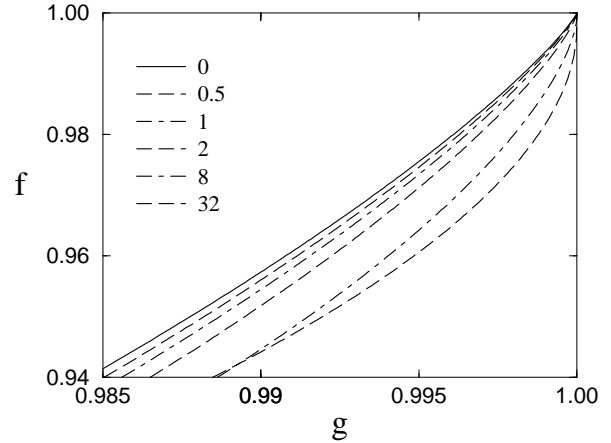}\\
        \end{center}

\caption{Plots of the leading $(1-g) \rightarrow 0$ behavior
of $f$, the fractional peak splitting, as a function of $g$,
the dimensionless interdot channel conductance, for a two-channel
connection between two quantum dots ($N_{\text{ch}} = 2$).
Each curve corresponds to a different value of the quantity
$2 \pi U_{\rho}/\hbar \omega_{P}$ (see legend to the left of the
curves).  The solid line is the result for an interdot barrier
that has effectively zero width
($2 \pi U_{\rho}/\hbar \omega_{P} = 0$).  The dashed and
dot-dashed curves show the leading $f$-versus-$g$ dependence for
finite barriers with $2 \pi U_{\rho}/\hbar \omega_{P}$ taking
on values from $0.5$ to $32$.   The curves are expected to be
substantially accurate at least for $(1-g)$ greater than or
equal to approximately $10^{-3}$ and for values of 
$2 \pi U_{\rho}/\hbar \omega_{P}$ not much larger than $2$.}
\label{fig:3}
\end{figure}

As for the single-channel system, one might inquire about
the nature of the leading analytic behavior of these 
corrections.  The answer is essentially parallel,
both in reasoning and substance, to that found at the
end of Sec.~III.  The constraints on $(1-g)$ are the same 
(i.e., 
$10^{-3} \lesssim (1-g) \ll e^{-(2/\pi)(2\pi U_{\rho}/\omega_{P})}$), 
and the leading finite-barrier corrections
to $Z(\epsilon_{F}, U_{\rho}, \omega_{P}, \rho)$ are expected to be 
roughly proportional to
$(2 \pi U_{\rho}/\hbar \omega_{P}) /|\ln (1-g)|$
when $2 \pi U_{\rho}/\hbar \omega_{P} \lesssim 2$.
In other words, $Z(\epsilon_{F}, U_{\rho}, \omega_{P}, \rho)$ 
can be expanded as follows:
\begin{eqnarray}
\frac{Z(\epsilon_{F}, U_{\rho}, \omega_{P}, \rho)}
        {\cos (\pi \rho/2) %
         \sqrt{2 e^{\gamma} \hbar v_{F} U_{\rho} / \pi^3}}
 & = & \sqrt{1-g} 
        \nonumber \\
 & & \hspace{-0.2in} + \, \, 
        c_{2} \frac{2 \pi U_{\rho}/\hbar \omega_{P} }{|\ln (1-g)|}
        \, \, + \, \ldots \, ,
\label{eq:Zfin}
\end{eqnarray}
where $c_{2}$ is a small positive number. 
Because, from prior work,~\cite{Matveev34,%
Golden1,Golden2} the leading corrections to the fractional 
peak splitting are proportional to 
$Z(\epsilon_{F}, U_{\rho}, \omega_{P}, \rho)^{2} 
\ln Z(\epsilon_{F}, U_{\rho}, \omega_{P}, \rho)$ at $\rho = 0$,
the leading finite-barrier correction to the fractional peak
splitting is roughly proportional
to $(2 \pi U_{\rho}/\hbar \omega_{P}) \, \sqrt{1-g}
\, \, \{ 1 - |\ln{(1-g)}|^{-1} \}$.
The corrections shown in Figure~3 follow this predicted
behavior quite well: for 
$2 \pi U_{\rho}/\hbar \omega_{P} \leq 2$, the 
analytic prediction captures the calculated corrections
with error margins of about 10\% or less for the $(1-g)$-dependence
and of about 20\% or less for the 
$2 \pi U_{\rho}/\hbar \omega_{P}$-dependence.
The numerical results suggest that
$c_{2}$ is slightly less than $0.02$.

\section{Conclusion}

This paper shows that
bosonization techniques for studying the behavior of
one-dimensional systems need not be abandoned when 
finite-length barriers (or constrictions) are introduced.
The finite-length effects of those barriers can be captured
to leading order by a standard perturbative approach, albeit
one that requires complicated calculations
and a fair degree of care.  Thus,
bosonization techniques can still be useful when nontrivial
behavior in the barrier region is at issue, and the 
approach presented in this paper may help 
investigators to distinguish between single-particle and
many-particle effects from a barrier's finite length.

With regard to the more particular problem of the 
Coulomb blockade behavior of coupled quantum dots,
our results can be summarized as follows.
This paper shows that at least for
one-channel or two-channel systems, the fractional peak
splitting $f$ of two strongly coupled dots decreases,
for a given value of the interdot conductance,
as the ratio $2 \pi U_{\rho}/\hbar \omega_{P}$ is
increased.  For one-channel systems ($N_{\text{ch}} = 1$),
the downward correction behaves as 
$(2 \pi U_{\rho}/\hbar \omega_{P}) /|\ln (1-g)|$ in
the limit where $2 \pi U_{\rho}/\hbar \omega_{P} \lesssim 2$
and $(1-g) \ll 1$.  Thus, for $N_{\text{ch}} = 1$, the
fractional peak splitting has the following leading-order
functional form: 
\begin{equation}
f_{1} = 1 - c_{1,1} \sqrt{1-g} 
        - c_{1,2} 
  \frac{(2 \pi U_{\rho}/\hbar \omega_{P})}{|\ln(1-g)|} \, ,
\label{eq:f1approx}
\end{equation}
where $c_{1,2}$ is somewhat less than 0.05 and, 
as derived in prior work,~\cite{Matveev34,Golden1,Golden2} 
$c_{1,1} = 8 e^{\gamma}/\pi^2$ (about $1.44$).
In two-channel systems ($N_{\text{ch}} = 2$),
and in the same limits,
the downward correction due to the finite barrier width
behaves as $(2 \pi U_{\rho}/\hbar \omega_{P}) \, \sqrt{1-g}
\, \, \{ 1 - |\ln{(1-g)}|^{-1} \}$,
resulting in leading-order behavior of the form
\begin{eqnarray}
f_{2} & = & 1 - c_{2,1} (1-g)|\ln(1-g)|
        \nonumber \\
 & & \hspace{-0.1in} \, 
        - \, c_{2,2} \,
        (2 \pi U_{\rho}/\hbar \omega_{P}) \, \sqrt{1-g} 
        \, \, \left\{ 1 - \frac{1}{|\ln{(1-g)}|} \right\} \, ,
\label{eq:f2approx}
\end{eqnarray}
where $c_{2,2}$ is somewhat less than $0.04$ and, as derived
in prior work,~\cite{Matveev34,%
Golden1,Golden2} $c_{2,1} = 16 e^{\gamma}/\pi^3$ (or about
$0.919$).  (It should be noted that prior results indicate
that, in the $N_{\text{ch}} = 2$ equation, the next additive
term independent of $2 \pi U_{\rho}/\hbar \omega_{P}$
will equal $-c_{2,3} (1-g)$, where 
$c_{2,3} \simeq 0.425$.~\cite{Golden2}  Because this term
linear in $(1-g)$ comes from higher-order terms in the 
effective action than those we have considered here
(see, e.g., Eq.~\ref{eq:SpinActfin}), we have omitted it
in our calculations and in the graphs of Figure~3 in order
to have a truer comparison between the 
$2 \pi U_{\rho}/\hbar \omega_{P}$-dependent terms and 
those they directly correct.
If need be, the linear term is easy enough to
combine with the result expressed in 
Eq.~\ref{eq:f2approx}.)

The above results combine with an earlier study of
the weak-coupling regime ($g \ll 1$)~\cite{Golden3}
to give a more complete understanding of the
$f$-versus-$g$ curve when one leaves the 
delta-function barrier
limit $2 \pi U_{\rho}/\hbar \omega_{P} \rightarrow 0$.
In particular, we come to the nontrivial conclusion 
that, for the experimentally realized values of
$2 \pi U_{\rho}/\hbar \omega_{P}
\simeq 1$,~\cite{Waugh,Crouch,Livermore,Golden3}
the corrections to the
results derived from modeling the barrier as
a delta function are not fundamentally substantial; 
thus, to this extent at least,
earlier theoretical work using a delta-function 
potential was correct.

In addition to confirming the essential nature of
the $f$-versus-$g$ curve for 
$2 \pi U_{\rho}/\hbar \omega_{P} \lesssim 2$, this
paper has helped us gain a better picture of what
happens to the $f$-versus-$g$ curve for more
general values of $2 \pi U_{\rho}/\hbar \omega_{P}$.
The weak-coupling results suggested that, as
$2 \pi U_{\rho}/\hbar \omega_{P}$ is increased,
the $f$-versus-$g$ curve shifts upward for small
values of $g$ (i.e., for $g \ll 1$) and
becomes flatter for intermediate
values of $g$.  The strong-coupling results suggest
(as conjectured) that, as the same ratio is increased, the
$f$-versus-$g$ curve shifts downward for large
values of $g$ (i.e., for $(1-g) \ll 1$) and becomes
flatter for intermediate values of $g$.
Together, the two sets of results
suggest that, in the limit of an extremely wide,
``adiabatic'' barrier 
($2 \pi U_{\rho}/\hbar \omega_{P} \rightarrow \infty$), 
the $f$-versus-$g$ curve
will be essentially flat for values of $g$
in an intermediate region between $0$
and $1$.  For such an adiabatic barrier,
the $f$-versus-$g$ curve will sit at some
essentially constant value of $f$ for most of this
interval and will turn sharply toward the
limiting values of $f = 0$ and $f = 1$ at the
edges.

For the moment, the intermediate value at which
the ``adiabatic'' $f$-versus-$g$ curve sits remains
a mystery. The weak-coupling results displayed an
antisymmetry around $g = 1/2$
that, if true at higher
orders, might aid in solving
for the fractional peak splitting when $g$ takes
on intermediate values.~\cite{Golden3}
Unfortunately, the strong-coupling results do not
exhibit such a simple symmetry.
As Figures~1 and 2 indicate, unlike the weak-coupling
results, the strong-coupling results do not reveal a
common point of intersection for
the leading-order curves that correspond to different
values of $2 \pi U_{\rho}/\hbar \omega_{P}$.
The apparent lack of a common pivot will
presumably make solution of the intermediate-$g$
problem more difficult.

\section*{Acknowledgments}
This work was supported by NSF grants No.
DMR99-81283 and No. DMR98-09363, and DARPA
grant No. 1034315.  The authors
thank R. Westervelt, C.H. Crouch, and
C. Livermore for helpful conversations.

\end{multicols}


\begin{references}

\bibitem{Waugh} F. R. Waugh, M. J. Berry, D. J. Mar, R. M. Westervelt,
  K. L. Campman, and A.~C. Gossard, Phys. Rev. Lett.
  {\bf 75}, 705 (1995); F.~R.~Waugh, M. J. Berry,
  C.~H. Crouch, C.~Livermore, D. J. Mar, R. M. Westervelt,
  K. L. Campman, and A.~C. Gossard, Phys. Rev. B {\bf 53}, 1413 (1996);
  F. R. Waugh, Ph.D. thesis, Harvard University, 1994.

\bibitem{Crouch} C. H. Crouch, C. Livermore, F. R. Waugh,
  R.~M.~Westervelt, K.~L.~Campman, and
  A.~C. Gossard, Surf. Sci. {\bf 361-362}, 631 (1996).

\bibitem{Livermore} C. Livermore, C. H. Crouch, R. M. Westervelt,
  K. L. Campman, and A. C. Gossard,
  Science {\bf 274}, 1332 (1996).

\bibitem{Flensberg} K. Flensberg, Physica B {\bf 203}, 432 (1994);
  Phys. Rev. B {\bf 48}, 11~156 (1993).

\bibitem{Matveev2} K. A. Matveev, Phys. Rev. B {\bf 51}, 1743 (1995).

\bibitem{Molen} L. W. Molenkamp, K. Flensberg, and M. Kemerink,
  Phys. Rev. Lett. {\bf 75}, 4282 (1995).

\bibitem{Matveev34} K. A. Matveev, L. I. Glazman, and H. U. Baranger,
  Phys. Rev. B {\bf 53}, 1034 (1996);
  {\bf 54}, 5637 (1996).

\bibitem{Golden1} J. M. Golden and B. I. Halperin, Phys. Rev. B
  {\bf 53}, 3893 (1996).

\bibitem{Golden2} J. M. Golden and B. I. Halperin, Phys. Rev. B
  {\bf 54}, 16 757 (1996).

\bibitem{Golden3} J. M. Golden and B. I. Halperin, Phys. Rev. B {\bf 56},
4716 (1997).

\bibitem{Liu}  Y.-L. Liu, Phys. Rev. B {\bf 56}, 6732
(1997).  Unfortunately, we have not been able to follow the 
reasoning of Liu's paper, which reports results that contrast 
sharply with our results here and with those derived in Refs.~7-10. 

\bibitem{Note1} It should be noted that the two-dot system that 
we have described has a close mathematical relation to other 
systems, including one consisting
of a single dot coupled to one lead through a nearly open point
contact.  See Refs. 7 and 8 above.   Information about the energy 
of this single dot can be obtained by measuring the capacitance 
between the dot and a nearby overlying gate.  See Ref. 5 above.

\bibitem{Thomas} K. J. Thomas, J. T. Nicholls, M. Pepper, W. R.
Tribe, M. Y. Simmons, and D. A. Ritchie, Phys. Rev. B {\bf 61},
R13,365 (2000); K. J. Thomas, J. T. Nicholls, N. J. Appleyard,
M. Y. Simmons, M. Pepper, D. R. Mace, W. R. Tribe, and D. A.
Ritchie, Phys. Rev. B {\bf 58}, 4846 (1998).

\bibitem{Emery} V. J. Emery, ``Theory of the One-Dimensional
Electron Gas,'' in {\it Highly Conducting One-Dimensional Solids},
edited by J. T. Devreese, R. P. Evrard, and V. E. van Doren
(Plenum, New York, 1979), pp. 247-303.

\bibitem{Heidenreich} R. Heidenreich, R. Seiler, and
D. A. Uhlenbrock, J. Stat. Phys. {\bf 22}, 27 (1980).

\bibitem{Haldane} F. D. M. Haldane, Phys. Rev. Lett. {\bf 47},
1840 (1981); J. Phys. C {\bf 14}, 2585 (1981).

\bibitem{Fradkin} E. Fradkin, {\it Field Theories of Condensed
Matter Systems}
(Addison-Wesley, Reading, 1991), pp. 74-88.

\bibitem{Schulz} H. J. Schulz, in
{\it Strongly Correlated Electronic Materials: The Los
Alamos Symposium, 1993}, edited by K. Bedell et al.
(Addison-Wesley, Reading, Massachusetts, 1994), pp. 187ff.;
in {\it Mesoscopic Quantum Physics, Les Houches,
Session LXI, 1994}, edited by E.~Akkermans, G. Montambaux,
J.-L. Pichard, and J. Zinn-Justin (Elsevier, Amsterdam, 1995),
pp. 533ff.


\bibitem{Shankar} R. Shankar, Acta Phys. Pol. B {\bf 26},
1835 (1996).

\bibitem{Kane} C. L. Kane and M. P. A. Fisher, Phys. Rev. Lett.
  {\bf 68}, 1220 (1992); Phys. Rev. B {\bf 46}, 7268 (1992);
  {\bf 46}, 15 233 (1992).

\bibitem{Lal} S. Lal, S. Rao, and D. Sen, Phys. Rev. Lett. 
  {\bf 87}, 026801 (2001).

\bibitem{Buttiker} M. B\"{u}ttiker, Phys. Rev. B {\bf 41}, 7906
  (1990).




\bibitem{ExplanFunc} Because we are only concerned with
the movement of electrons through the channel between the
dots (rather than motion within the dots themselves), 
we use free boundary conditions, and the 
eigenfunctions that result consist of 
incident, transmitted, and
reflected parts. 

\bibitem{Oops} J. N. L. Connor, Mol. Phys. 
{\bf 15}, 37 (1968). 

\bibitem{Miller} J. C. P. Miller in {\it Handbook of
Mathematical Functions}, edited by M. Abramowitz and
I. A. Stegun (National Bureau of Standards, Washington,
D.C., 1964), Appl. Math. Ser. 55, p. 685.


\end{references}
\end{document}